\documentclass[prb,amssymb,amsmath,twocolumn,showpacs]{revtex4}
\usepackage{bm}
\usepackage{graphicx}
\usepackage{color}

\begin{document}

\title{High magnetic field theory for the local density of states in graphene with
smooth arbitrary potential landscapes}

\author{Thierry Champel}
\affiliation{Laboratoire de Physique et Mod\'{e}lisation des Milieux
Condens\'{e}s, CNRS and Universit\'{e} Joseph Fourier, B.P. 166, 25 Avenue des
Martyrs, 38042 Grenoble Cedex 9, France}

\author{Serge Florens}
\affiliation{Institut N\'{e}el, CNRS and Universit\'{e} Joseph Fourier, B.P.
166, 25 Avenue des Martyrs, 38042 Grenoble Cedex 9, France}

\date{\today}

\begin{abstract}
We study theoretically the energy and spatially resolved local density of states
(LDoS) in graphene at high perpendicular magnetic field.
For this purpose, we extend from the Schr\"odinger to the Dirac case a semicoherent-state
Green's-function formalism, devised to obtain in a quantitative way the lifting of the Landau-level
degeneracy in the presence of smooth confinement and smooth disordered potentials.
Our general technique, which rigorously describes quantum-mechanical motion in a magnetic field beyond 
the semi-classical guiding center picture of vanishing magnetic length (both for the ordinary two-dimensional electron
gas and graphene), is connected to the deformation (Weyl) quantization theory in phase space developed 
in mathematical physics.
For generic quadratic potentials of either scalar (i.e., electrostatic) or mass
(i.e., associated with coupling to the substrate) types, we exactly solve the regime of large magnetic field  (yet at finite magnetic length - formally, this amounts to considering an infinite Fermi velocity) where Landau-level mixing becomes negligible.
Hence, we obtain a closed-form expression for the graphene Green's function in this regime, 
providing analytically the discrete energy spectra for both cases of scalar and
mass parabolic confinement.
Furthermore, the coherent-state representation is shown to display a hierarchy of local energy 
scales ordered by powers of the magnetic length and successive spatial derivatives of the local 
potential, which allows one to devise controlled approximation schemes at finite temperature 
for {\it arbitrary} and possibly disordered potential landscapes.
As an application, we derive general analytical non-perturbative expressions for the LDoS, which 
may serve as a good starting point for interpreting experimental studies.
For instance, we are able to account for many puzzling features of the LDoS
recently observed by high magnetic field scanning tunneling spectroscopy
experiments on graphene, such as a roughly $\sqrt{m}$-increase in
the $m$th Landau-level linewidth in the LDoS peaks at low temperatures, 
together with a flattening of the spatial variations in the Landau-level
effective energies at increasing $m$.

\end{abstract}


\maketitle

\section{Introduction}

\subsection{Quantum-Hall effect in graphene}

The observation of an anomalous quantization of the Hall resistance in graphene
at high magnetic fields,  \cite{Novo2005,Zhang2005,Novo2007} related to the
massless, relativistic-like spectrum of low-energy electrons on the
two-dimensional honeycomb lattice, has triggered much excitation in recent
years, see Ref. \onlinecite{Castro2009} for a review. Indeed, the experimentally
measured Hall resistance follows the Landau-level structure expected for massless 
Dirac electrons, 
\cite{Zheng2002,Gusynin2005} $E_{m}=\pm \sqrt{m} \hbar \Omega_{c}$ in the clean
case, with $m$ a positive integer and $\Omega_{c}=\sqrt{2} v_{F}/l_{B}$ the
graphene characteristic frequency given in terms of the Fermi velocity $v_{F}$ and
of the magnetic length $l_{B}=\sqrt{\hbar c/|e|B}$ (here $e=-|e|$ is the
electron charge, $c$ the speed of light, and $B$ the magnetic field strength).
The $\sqrt{B}$ dependence of the characteristic frequency $\Omega_c$ in graphene, to be
contrasted with the linear dependence of the cyclotron frequency
$\omega_{c}=|e|B/(m^{\ast}c)$ of more standard two-dimensional electron gases (2DEGs)
based on semiconducting heterostructures (in this case, $m^{\ast}$ is the
electronic effective mass) described by Schr\"{o}dinger equation,
constitutes one of the main signatures used so far in experiments to exhibit the
relativistic-like character of the massless charge carriers.

Also quite remarkable is that graphene displays a surface opened to the outside world,
providing a direct window to its electronic excitations. This is a clear experimental
advantage of graphene compared to 2DEGs
based on semiconducting heterostructures, where the 2DEG is buried deep inside the structure
(typically 100 nm or more).
Graphene thus offers the opportunity to obtain precise insights into local physical properties
of quantum-Hall systems, such as the local density of states (LDoS) via scanning
tunneling spectroscopy (STS) measurements. In contrast, such local probes
experiments have very poor spatial resolution in ordinary heterostructures, although
some progress has been made recently, see Ref. \onlinecite{Hash2008}.
This technical advantage will be certainly important in the future to elucidate the relation
between microscopic inhomogeneities induced by various disorder types and macroscopic
transport properties of large samples. Various open questions in this respect are
the nature of the universal plateau to plateau quantum phase transition, \cite{Ludwig1994,Goswami2007,Giesbers2009} or on a
more quantitative level the precise formation of wide Hall plateaus.
To pursue this goal, STS is one of the interesting available experimental techniques, and
first experiments in graphene at high magnetic field have been performed recently. \cite{Li2009,Miller2009}
Since this spectroscopic method gives direct information on the local electronic states,
a better understanding of the LDoS, specific to the case of graphene at high magnetic fields
and in arbitrary potential landscapes (without proceeding to disorder averaging), needs to be
achieved. This is the main aim of the present paper. A second important
aspect of our work is to obtain analytical solutions for a large class of parabolic 
confinement models, and as a motivation we now discuss the different
types of potentials that can be involved in the two-dimensional Dirac
Hamiltonian.

\subsection{Disorder types for graphene}

Because of the multicomponent structure of the wave function for graphene, several
types of disorder can occur, which we introduce here.
The quasiparticle dispersion for graphene has two Dirac cones (two ``valleys'')
at low energies. For a given valley, the Hamiltonian in the presence of a
perpendicular magnetic field has a matrix structure and is written as
\begin{equation}
H_{0}= v_{F} {\bm \sigma} \cdot \hat{\bm \Pi},
\label{Hgraphene}
\end{equation}
where $v_{F}$ is the Fermi velocity, ${\bm \sigma}$ is a vector whose components
are the Pauli matrices $\sigma_{x}$ and $\sigma_{y}$ in the pseudospin space, and
the momentum operator is
\begin{equation}
\hat{{\bm \Pi} }=-i \hbar {\bm \nabla}_{{\bf r}}-\frac{e}{c}{\bf A}({\bf r}).
\end{equation}
The vector potential ${\bf A}$ is related to the uniform transverse magnetic field ${\bf B}$ via
the relation ${\bm \nabla} \times {\bf A}={\bf B}=B \hat{{\bf z}}$.
For convenience, we will omit both physical spin and valley indices, thus assuming
that the two valleys of graphene remain completely decoupled from each other and
can be studied separately. \cite{Castro2009}

Quite generally, potential terms appear as either a random scalar potential,
a random Dirac mass or a random vector potential. \cite{Ludwig1994} The Hamiltonian in presence
of these potentials is given by
\begin{equation}
H=H_{0}+ V({\bf r}), \label{prob}
\end{equation}
where the function $V({\bf r})$ takes the general form
\begin{equation}
 V({\bf r}) =\sum_{p=s,x,y,z} \sigma_{p} V_{p}({\bf r})
\label{prob2}
\end{equation}
with $\sigma_{s}$ the identity matrix in the pseudospin space, associated to
the scalar potential term $V_{s}({\bf r})$. This contribution may have many different
physical origins: electrostatic confinement potential, impurity random potential,
and/or Hartree potential resulting from the mean-field mutual Coulomb interaction
between the electrons.
The diagonal but antisymmetric term $V_{z}({\bf r})$, associated to the
$\sigma_{z}$ Pauli matrix, describes the so-called random mass potential.
This contribution might be introduced by the underlying substrate in
single-layer graphene, while in bilayer graphene, such a term can be produced in a
controllable way by introducing different electrostatic potentials in the two
layers. \cite{Ohta2006} The off-diagonal contributions coming as ${\bf V}({\bf r})=\left[V_{x}({\bf r}),
V_{y}({\bf r})\right]$ can be associated with a random vector potential, coming
from the spatial distortion of the graphene sheet in the third dimension by ripples. \cite{Morozov2006,Castro2009}
In what follows, all three possible types of disorder will be considered within the high 
magnetic field regime.

\subsection{Existing theoretical results for graphene in various potential types}

Let us first discuss various toy models of potentials (in a magnetic field)
that were studied in the recent graphene literature.
Quite generally, within the Dirac equation fewer models can be solved exactly than
within its non-relativistic counterpart. For instance, the classic one-dimensional parabolic
confinement model, as well as the circular parabolic confinement model, are
seemingly not analytically tractable. For the 2DEG, the former is the well-known model to
introduce the edge states and explain the quantized conductance in Hall bars.
The latter is the basic model for quantum dots and leads under magnetic field
to the Fock-Darwin states with discrete energies.
Thus, only much simpler models can be solved analytically for graphene, such as the uniform
electric field. \cite{Lukose2007,Peres2007} Progress can be achieved for circular hard-wall
confinement with either scalar \cite{Recher2008} or mass \cite{Schnez2008}
potentials, but only a solution in terms of special functions is then possible.
For parabolic and more complex potentials, fully numerical methods have to be used, e.g.,
see Ref. \onlinecite{Chen2007}.
We will show in this paper that the limit of negligible Landau level mixing allows one to solve
analytically a large class of parabolic models, providing new insights in the high magnetic
field regime.

Coming to the more complex question of disorder, even less is actually known.
Recent work devoted to the quantum-Hall effect in graphene has proposed to take into account disorder phenomenologically in the
expression of Green's function by adding a constant imaginary part $i \Gamma$ in
the self-energy, \cite{Gusynin2005,Gusynin2006} but Hall quantization obtains only
in the limit where the energy rate $\Gamma \rightarrow 0$.
The LDoS  in the vicinity of a single pointlike impurity   and in the presence of a strong magnetic field has been 
studied recently. \cite{Biswas2009,Bena2009}
Various types of disorders were also considered in Refs. \onlinecite{Peres2006} and \onlinecite{Ostrovsky2008}
within the self-consistent Born approximation. While this method may be
justified for short-range scatterers, it turns out \cite{Raikh1993} to be inappropriate 
for a smooth potential in high magnetic fields. Because a quasi-local picture
takes place in the high magnetic field regime, \cite{Champel2009} our calculation will be able to provide 
accurate expression for the LDoS in smooth {\it arbitrary} potentials.

\subsection{High magnetic field regime}

The strategy to follow is best explained by starting to discuss the specific nature of
disorder for 2DEGs at high magnetic field.
For very clean heterostructures, the disordered potential seen by the electrons
is mostly smooth on large length scales (several tens of nanometers), as the majority of
impurities sit far away from the 2DEG. In contrast to the low magnetic field regime, where
the electrons explore ergodically macroscopic regions of the sample, the high
field regime is characterized by cyclotron motion close to equipotential lines
of potential landscape $V(\bf r)$ with a narrow transverse spread proportional to the magnetic
length (which is smaller than 10 nm at several tesla). The disorder landscape felt by the
electronic wave functions is therefore very smooth in that situation.
We note that in graphene additional sharper potential variations (such as atomic vacancies
of the carbon layer, or local imperfections from the nearby substrate) can occur, although these
tend to be detrimental to quantum-Hall physics by increasing the mixing of Landau levels. The
coupling to the substrate can however be removed by suspending graphene flakes
or with a decoupled layer in epitaxial graphene, \cite{Miller2009} resulting in very high
mobility samples. In fact, for both non-relativistic 2DEGs and graphene, the essence of the quantum
Hall effect lies already by considering smooth potential variations only, which is the case to be
followed from now on.

Theoretically, this smooth disorder regime was shown to be problematic at high magnetic field
for standard quantum-mechanical methods based on perturbative expansions in potential strength. \cite{Raikh1993}
In that case, the high magnetic field regime is the correct starting point, and is characterized by
two different dimensionless small parameters:
(i) $l_B/\xi$ associated to the transverse spread of the wave function along the
classical guiding center ${\bf R}$, with $l_B$ the magnetic length 
and $\xi$ the typical length scale related to local variations in the potential;
(ii) $l_B |{\bm \nabla} V|/\hbar \omega_c \simeq l_B \delta V/ \xi \hbar \omega_c$ associated 
to Landau level mixing by local gradients $|{\bm \nabla}V|$ of the potential, introducing 
$\delta V$ the typical amplitude variations in the potential on the scale $\xi$, and the cyclotron
frequency $\omega_c$ in the 2DEG case.

Clearly, quantum mechanics calls for non-zero $l_B/\xi$, otherwise the so-called
semiclassical guiding center picture at $l_B=0$ emerges, giving at best a
qualitative picture, and missing important quantum effects such as level
quantization, tunneling, or interferences effects due to the potential energy $V({\bf R})$.
The second parameter $l_B |{\bm \nabla} V|/\hbar \omega_c$ controls the degree
of Landau level mixing, so that Landau levels strictly decouple at infinite $\omega_c$.
Most previous works have considered either limits separately (either $l_B\to 0$
or $\omega_c\to \infty$), and the necessary formalism to incorporate both
non-zero $l_B$ and finite $\omega_c$ was developed for the standard
2DEG by the authors in Refs. \onlinecite{Champel2007,Champel2008,Champel2009}, which will be extended in
the present paper to the case of graphene. This mathematical construction shows that a local picture
of the high magnetic field physics emerges in terms of semicoherent-state Green's function,
with a hierarchy of local energy scales \cite{Champel2009} ordered by powers of the magnetic
length and successive spatial derivatives of the confinement or disordered potential.

In the simplified, yet fully quantum limit of infinite cyclotron frequency and
non-zero $l_B$, initial progress was made by other authors in 
Refs.~\onlinecite{Girvin} and \onlinecite{Jain} for the 2DEG case, where it was shown that
Schr\"odinger equation acquires a uni-dimensional character, offering an
analysis for toy models of confinement or tunneling in the lowest Landau level.
The general structure of this limit was clarified in further developments in the
Green's-function formalism, \cite{Champel2009, Champel2009bis} and this will be
also examined in detail for graphene in the present paper. Our methodology is
based on the exclusive use of Green's functions, not wave functions, for the
simple reason that we project the quantum dynamics onto a semi-coherent
representation with nonorthogonal states, forcing us {\em de facto} to give up
the wave-functions picture. Noticeably, because the overcomplete character of
the chosen representation allows one to get rid of the Hilbert-space formulation
inherent to the traditional operator formulation of quantum mechanics, a
unification of closed and open systems quantum mechanics is made possible
here, i.e., one can get and treat quantization and lifetime effects on an equal
footing. An important application is the possibility to write down in the 2DEG
case a unique Green's-function expression which holds for all cases of quadratic
potentials. This derivation has clearly proved that the appearance of lifetimes
(expressing the presence of decaying states, i.e., an intrinsic time asymmetry)
has for physical origin the instability of the dynamics occurring at saddle
points of the potential landscape; see Ref. \onlinecite{Champel2009} for a
thorough discussion of this point.

In the graphene case, our calculation at large characteristic frequency $\Omega_{c}$ (or equivalently at large Fermi velocity)
brings important information, because, in contrast to standard 2DEGs, even
simple models of parabolic confinement for graphene do not possess an analytic
solution at finite $\Omega_c$. However, we will show that
the limit $\Omega_c\rightarrow+\infty$ is exactly solvable for most quadratic potentials,
allowing us to extract the explicit discrete energy spectrum in case of
several parabolic confinement models, and also, in principle, the transmission coefficients in
case of tunneling near saddle points. Going beyond these toy models, our general formalism also
allows us to calculate in a controlled way the LDoS in an {\it arbitrary} and
possibly disordered potential landscape. Our results will be discussed with respect to recent 
experimental findings. \cite{Li2009,Miller2009}

\subsection{Structure of the paper and summary of results}

First, in Sec.~\ref{sec:free}, we shall investigate the free Dirac Hamiltonian
in a transverse magnetic field, and introduce the graphene vortex states, which are
the building blocks of the whole theory developed here. These states form an
overcomplete family of semicoherent states, strongly localized around arbitrary
guiding center positions ${\bf R}$, and encode the cyclotron motion quantum mechanically.

In subsequent Sec.~\ref{sec:formalism}, we introduce the Green's function for
graphene vortex states and derive its general equation of motion,
Eq.~(\ref{Dysonfinal}), including Landau-level mixing processes. The general connection to
the real-space Green's function is also explicitly made in Eq.~(\ref{passage2}),
allowing one to calculate, in principle, any physical observable.

In Sec.~\ref{sec:high}, we show that the problem simplifies greatly in the
limit of negligible Landau level mixing. First, for locally flat potentials (away from
saddle points or bottom of potential wells), we find that the $m$th Landau
level acquires a dependence on the position ${\bf R}$, according to the simple formula,
\begin{eqnarray}
\xi_{m,\pm}({\bf R})= \tilde{v}_{m}^{+}({\bf R}) \pm
\sqrt{(\hbar \Omega_c \sqrt{m})^{2}+\left[\tilde{v}_{m}^{-}({\bf R}) \right]^{2}}.
\label{effpot}
\end{eqnarray}
Here $\tilde{v}_{m}^{+}({\bf R})$ and $\tilde{v}_{m}^{-}({\bf R})$ are renormalized effective 
potentials that are simple functionals of the bare scalar and mass potentials ; for their definitions  in terms of $V_{s}$ and $V_{z}$,
see Eq. (\ref{tildev}) and the associated discussion in Sec.~\ref{Matrix}.
Second, when curvature of the potential is included, we find that simple analytic
solutions for several parabolic models can be obtained.
In particular, for circular parabolic {\it scalar} potential $V_s({\bf r}) =
(1/2) U_0 (x^2+y^2)$, the discrete energy spectrum (in terms of Landau level
index $m \geq 1$ and an extra quantum number $n$, which is a positive
integer $\geq 1$) reads:
\begin{equation}
E_{m,n} = \pm \hbar \Omega_c \sqrt{m} + l_B^2 U_0 (m+n+1/2)
\label{spectrum1}
\end{equation}
(we have assumed $\Omega_c\gg l_B^2 U_0$ above).
Apart from the well-known anomalous Landau-level quantization with respect to
quantum number $m$, this result is very reminiscent of Fock-Darwin states for
standard 2DEGs with respect to the linear dependence in the integer $n$.
More interestingly, for circular parabolic {\it mass} potential $V_z({\bf r}) =
(1/2) U_0 (x^2+y^2)$, the discrete energy spectrum displays now an anharmonic
form
\begin{equation}
E_{m,n} = -\frac{l_{B}^{2}U_{0}}{2} \pm \sqrt{(\hbar \Omega_c \sqrt{m})^2
+ (l_B^2 U_0 [m+n+1/2])^2}
\label{spectrum2}
\end{equation}
that was not obtained to our knowledge.
Generalization to non-circularly symmetric parabolic potentials is also readily
obtained, as well as for the combination of uniform scalar and parabolic mass terms
(and vice versa), with detailed calculations appearing in several appendices.
However, we show that potentials that combine sizeable spatial variations in
both scalar and mass terms are in general not analytically tractable, even in the high
magnetic field regime, except for the lowest Landau level.

In Sec.~\ref{sec:deformation}, we make explicit the connection of our formalism to the
so-called deformation (or Weyl) quantization, which corresponds to the proper way of
quantizing the dynamics in phase space. For two-dimensional problems in a magnetic field, the
vortex state formalism is in fact performing a mixed representation of phase
space in terms of the two-dimensional coordinates of the center of mass together
with a discrete quantum number associated to Landau levels while standard Weyl
quantization would introduce a four-dimensional description in terms of positions
and momenta of the electron. This latter choice is however unpractical for the
high magnetic field regime and this shows that the vortex states are most robust in this
regime. As should be expected, in the limit of infinite frequency ($\omega_{c} \to \infty$ for the 2DEG or $\Omega_{c} \to \infty$ for graphene), Landau
levels become fully decoupled, and the quantum dynamics reduces to a unidimensional
one in terms of the two vortex coordinates, acting as conjugate variables. An
effective one-dimensional picture of motion is thus rigorously obtained, overcoming
certain regularization problems of the path-integral technique. \cite{Jain}

Finally, in Sec.~\ref{sec:local}, we provide generic expressions for the LDoS
in an {\it arbitrary} scalar or mass potential that can be described {\it locally}
up to its first-order derivatives (generalized graphene drift states). Regarding recent
experimental findings, we show that:
\begin{itemize}
\item positions, amplitudes, and widths of the LDoS peaks qualitatively depend on the
dominant type (scalar or mass) of local potential, see, e.g., Eq.
(\ref{effective});
\item as the tip scans the surface, the LDoS peak energy of the
$m$th Landau level follows the effective potential given in
Eq.~(\ref{effpot}), see Fig.~\ref{figVeff}, so that the resulting energy variations 
{\it shrink} with increasing $m$, in agreement with the experimental findings
for graphene~\cite{Miller2009} and standard 2DEG;~\cite{Hash2008}
\item on the contrary, the width of the LDoS peaks at {\it fixed} 
tip position {\it grows} with increasing $m$ (roughly as $\sqrt{m}$), as observed 
in Ref. \onlinecite{Li2009} for graphene. Such a dependence is also expected for the ordinary 2DEG.
\end{itemize}

\section{Free Hamiltonian : Vortex states of graphene}
\label{sec:free}

\subsection{Vortex states for the standard 2DEG}

Before investigating the case of graphene under magnetic field, we briefly
recall the vortex states for the case of the non-relativistic 2DEG. This introduction will be useful to show that many physical
and technical aspects of the 2DEG can be directly transposed to the case of
graphene (studied in the next subsection).

A single free electron of effective mass $m^{\ast}$ confined in a $(xy)$
two-dimensional plane and subjected to a uniform magnetic field pointing in the
perpendicular direction ${\bf B}=B \hat{{\bf z}}$ is described by the
Hamiltonian
\begin{equation}
H_{2DEG}=\frac{\hat{{\bm \Pi}}^{2}}{2 m^{\ast}}=\frac{\hat{\Pi}_{x}^{2}+\hat{\Pi}_{y}^{2}}{2 m^{\ast}}.
\label{Ham2D}
\end{equation}
Then, the eigenvalue problem $H_{2DEG}\Psi=\varepsilon \Psi$ leads to the well-known
quantization of the kinetic energy into Landau levels,
\begin{equation}
\varepsilon_{m}=\left( m + \frac{1}{2} \right) \hbar \omega_{c}
\label{Sol1}
\end{equation}
with the cyclotron pulsation $\omega_{c}=|e|B/m^{\ast} c= \hbar
/m^{\ast}l_{B}^{2}$ and $m \geq 0$ a positive integer (here $l_{B}=\sqrt{\hbar
c/|e|B}$ is the magnetic length). It is important to note here the large
degeneracy of the Landau energy levels $\varepsilon_{m}$. Indeed, for the
motion of an electron in the two-dimensional plane, one expects at least two
quantum numbers since there are two degrees of freedom. The degeneracy means
that there is a great freedom in the choice of the second (degeneracy) quantum
number, or equivalently, in the choice of a basis of eigenstates $\Psi$.
Consequently, there exist in the literature different ways to derive the energy
quantization, Eq. (\ref{Sol1}). Eigenstates characterized by a peculiar symmetry of
the (gauge-invariant) probability density $|\Psi|^{2}$ are preferentially chosen
in many contexts. For instance, the Landau states, with a conserved momentum as the
degeneracy quantum number, are translationally invariant in one direction.
\cite{Landau} Circular eigenstates characterized by a rotation invariance
around the origin \cite{Circular} are also well known and often used. It is
worth stressing that the real difference between the Landau states and the
circular states is not the gauge because both kinds of states can be obtained
in any gauge. \cite{note} The real difference is in the choice of the
gauge-invariant quantum numbers, which are intimately related to the symmetry of
the probability density $|\Psi|^{2}$.

Importantly, both Landau and circular eigenstates do not reflect the symmetry
of the cyclotron motion around an arbitrary point ${\bf R}=(X,Y)$ in the $(x,y)$
plane so that the consideration of the classical limit with these sets of
states is rather tricky. Since they do not correspond to the classical picture
of the motion, it is difficult to appreciate the wave-particle duality.
By imposing that the probability density $|\Psi|^{2}$ of the eigenstates has the
same symmetry as the cyclotron motion, i.e., is a function of $\left|{\bf
r}-{\bf R}\right|$ only, we get \cite{Champel2007} the so-called vortex states,
given in the symmetrical gauge (${\bf A}=B \hat{{\bf z}} \times {\bf r}/2$) by
\begin{eqnarray}
\Psi_{m,{\bf R}}({\bf r}) =
\frac{1}{l_{B}\sqrt{2 \pi m!}}
\left[
\frac{x-X+i(y-Y)}{\sqrt{2}l_{B}}
\right]^{m} \nonumber \hspace*{1cm} \\
\times
\exp\left[
-\frac{(x-X)^{2}+(y-Y)^{2}+2i(yX-xY)}{4 l_{B}^{2}}
\right]
.
\label{Sol2}
\end{eqnarray}
For practical convenience, we shall now use the Dirac bracket notation by
writing $\Psi_{m,{\bf R}}({\bf r})=\langle {\bf r}| m,{\bf R} \rangle$.
Eigenstates (\ref{Sol2}) of Hamiltonian (\ref{Ham2D}), associated with energy
quantization (\ref{Sol1}), are characterized by the set of quantum numbers
$|m,{\bf R} \rangle$, where $m$ is a positive integer related to the
quantization of the circulation around the vortex and ${\bf R}=(X,Y)$ is a
continuous quantum number corresponding to the vortex location in the plane
[note with Eq. (\ref{Sol2}) the ``vortex''-like phase singularity at ${\bf
r}={\bf R}$ for $m \geq 1$, which justifies the chosen denomination for the set
of states]. These localized wave functions clearly encode the classical cyclotron
motion around the guiding center ${\bf R}$ quantum mechanically. The vortex states
form a semiorthogonal basis, with the overlap
\begin{eqnarray}
\langle m_{1}, {\bf R}_{1}|m_{2},{\bf R}_{2} \rangle
=
\delta_{m_{1},m_{2}}
\langle {\bf R}_{1}| {\bf R}_{2} \rangle, \label{overlap}
\end{eqnarray}
where
\begin{equation}
\langle {\bf R}_{1}| {\bf R}_{2} \rangle
=\exp\left[ -\frac{({\bf R}_{1}-{\bf R}_{2})^{2}-2i \hat{{\bf z}} \cdot ({\bf R}_{1} \times {\bf R}_{2})}{4 l_{B}^{2}}\right].
\label{overlapR}
\end{equation}
An important property is that the states (\ref{Sol2}) present the coherent
character with respect to the degeneracy quantum number ${\bf R}$, i.e., they
satisfy coherent states algebra. Note that these states are however eigenstates
of the free Hamiltonian associated to the Landau-level index $m$, and form
more precisely a semicoherent basis with respect to the quantum numbers
($m$,${\bf R})$. In particular, they also obey the following
completeness relation
\begin{eqnarray}
\int
 \!\!\! \frac{d^{2} {\bf R}}{2 \pi l_{B}^{2}} \sum_{m=0}^{+ \infty} | m,{\bf R} \rangle \langle m,{\bf R}
|
=1. \label{compvortex}
\end{eqnarray}
According to this relation (\ref{compvortex}) and general unicity properties
of the decomposition onto coherent states, \cite{Champel2007} it is possible to
expand arbitrary states or operators in the vortex state representation. Hence, despite
being nonorthogonal, the set of states $|m,{\bf R} \rangle$ with $m\geq 0$ does
form a basis of eigenstates, as the Landau and the circular states.

Besides providing a clear quantum mechanical dual of the classical cyclotron
motion, there are several good reasons to prefer specifically the vortex states
over an orthogonal set of eigenstates to study the process of lifting of the
Landau level degeneracy in the presence of a smooth {\em arbitrary} potential.
First, in contrast to the Landau states or circular states, the vortex states do
not impose a symmetry to the degeneracy quantum number, and thus permit a great
adaptability to the spatial variations in the local electric fields, coming from
either random impurity donors, confinement potentials, or macroscopic voltage
drops (in a nonequilibrium regime). This property leads to advantages in terms
of computability since it is possible in the vortex representation to calculate
and classify Landau-level mixing processes in a simple and natural manner (this
will be illustrated in Sec.~\ref{Matrix}). Second, at a more fundamental level, the
vortex states are expected to be quite insensitive to any kind of smooth
perturbations, since the quantum number $m$ has a purely topological origin in
the vortex representation (for the Landau states or circular states, the
quantization of the kinetic energy comes either partially or entirely from the
condition of vanishing of the wave function at infinity, what makes them much
less robust to perturbations as a result of their nonlocality). Owing to this
quantum robustness, the vortex states are thus naturally selected by the
dynamics in the presence of a smooth potential with an arbitrary spatial
dependence. They appear to be much more stable than their superpositions (for
instance, the Landau states) since they are the only states surviving under the
action of such an interaction potential without any internal symmetry.
Interestingly, the vortex states are also the best states to describe the
transition from quantum to classical. Despite being fully quantum, they thus
encode de facto classicality properties and insensitivity to openness of the
system. Therefore, they provide the best playground to understand the mechanisms
of irreversibility, decoherence and dissipation in high magnetic fields. We will
comment on this point in more detail later, in Sec.~\ref{hier}.

\subsection{Graphene vortex states}
We now come for good to graphene, which is described in the absence of potential by
Hamiltonian (\ref{Hgraphene}). By searching the wave functions under the
spinorial form
\begin{equation}
\tilde{\Psi} =\left(\begin{array}{c}
u \\w
\end{array}
 \right)
\end{equation}
with 
\begin{equation}
H_{0} \tilde{\Psi}=E \tilde{\Psi},
\end{equation}
we get the following equations:
\begin{eqnarray}
\left(\hat{\Pi}_{x}-i \hat{\Pi}_{y} \right) w &= & \frac{E}{v_{F}} u, \label{S1}
\\
\left(\hat{\Pi}_{x}+i \hat{\Pi}_{y} \right) u &= & \frac{E}{v_{F}} w \label{S2}
\end{eqnarray}
with $E$ the energy eigenvalue. Getting rid of the component $u$ we get the
Schr\"{o}dinger-type equation for the component $w$,
\begin{eqnarray}
\left(\hat{\Pi}_{x}+i \hat{\Pi}_{y} \right)\left(\hat{\Pi}_{x}-i \hat{\Pi}_{y}
\right) w &= & \left(\frac{E}{v_{F}}\right)^{2} w. \label{v}
\end{eqnarray}
Using that
\begin{equation}
\left[ \hat{\Pi}_{x}, \hat{\Pi}_{y}
\right]
= -i \hbar \frac{|e|B}{c}=-i \frac{\hbar^{2}}{ l_{B}^{2}},
\end{equation}
we find that Eq. (\ref{v}) reads
\begin{equation}
\hat{{\bm \Pi} }^{2}w
= \tilde{E} w  \label{E1}
\end{equation}
with
\begin{equation}
\tilde{E}= \left(\frac{E}{v_{F}} \right)^{2} + \frac{\hbar^{2}}{ l_{B}^{2}}.
\end{equation}
By posing $\tilde{E}=2 m^{\ast} \varepsilon$ in Eq. (\ref{E1}), where
$\varepsilon$ has the dimension of an energy, we directly recognize the
eigenproblem for a free 2DEG under magnetic fields discussed in the former
section. This mapping shows that there is also a great freedom to choose a
basis of eigenstates in the case of graphene. In the following, we introduce the
analog of vortex states, Eq. (\ref{Sol2}), for graphene.

 From Eq. (\ref{Sol1}), we directly deduce that
\begin{equation}
\tilde{E}=\left(2m+1\right) \hbar^{2} /l_{B}^{2}.
\end{equation}
Therefore, we get that the energy eigenvalues of the graphene Hamiltonian are
\begin{equation}
E_{m,\lambda}=\lambda \sqrt{m} \, \hbar \sqrt{2}\frac{v_{F}}{l_{B}}
=\lambda \sqrt{m} \, \hbar \Omega_{c}, \label{Em}
\end{equation}
where $\lambda$ is a band index, which is equal to $\pm 1$ if $m\geq 1$, and 0
if $m=0$. We see that the energy levels are no more equidistant in energy and
that the characteristic energy for graphene reads $\hbar \Omega_{c} \propto \sqrt{B}$
instead of $\hbar \omega_{c} \propto B$ for 2DEGs. The component $u$ of the spinorial
wave function $\tilde{\Psi}$ is straightforwardly obtained from the knowledge of
the component $w$ by using Eq. (\ref{S1}). The corresponding normalized
graphene vortex states are thus
\begin{equation}
\tilde{\Psi}_{m,{\bf R},\lambda}({\bf r})=\frac{1}{\sqrt{1+|\lambda|}} \left(
\begin{array}{c}
\lambda \Psi_{m-1,{\bf R}}({\bf r}) \\
 i \Psi_{m,{\bf R}}({\bf r})
\end{array}
 \right).
\end{equation}
 Within the Dirac notation, the set of vortex quantum numbers we shall consider
for graphene takes therefore the form
\begin{equation}
|m,{\bf R},\lambda \rangle
=
\frac{1}{\sqrt{1+|\lambda|}}
\left(
\begin{array}{c}
\lambda |m-1,{\bf R}\rangle
\\
i |m,{\bf R} \rangle
\end{array}
\right)
.
\label{Dirac}
\end{equation}
 The label $\lambda$ which characterizes the spinorial structure of the
eigenvectors appears here as an additional quantum number with respect to the
2DEG.

Using the semiorthogonality property, Eq. (\ref{overlap}), of the vortex states, we
can easily check that the graphene vortex states present the same property as
their ``non-relativistic'' counterparts. Indeed, we have
\begin{eqnarray}
\langle m_{1}, {\bf R}_{1},\lambda_{1} |m_{2},{\bf R}_{2},\lambda_{2} \rangle
= \frac{1}{\sqrt{1+|\lambda_{1}|}}\frac{1}{\sqrt{1+|\lambda_{2}|}} \hspace*{0.8cm} \nonumber \\
\times
\left(
\begin{array}{c}
\lambda_{1} \langle m_{1}-1,{\bf R}_{1} |
\\
- i \langle m_{1},{\bf R}_{1} |
\end{array}
 \right)
\cdot \left( \begin{array}{c}
\lambda_{2} |m_{2}-1,{\bf R}_{2}\rangle
\\
 i |m_{2},{\bf R}_{2}\rangle
\end{array}\right)
\nonumber
\\
= \delta_{m_{1},m_{2}} \langle {\bf R}_{1}| {\bf R}_{2}
\rangle
\left(
\frac{\lambda_{1}\lambda_{2}+1}{\sqrt{1+|\lambda_{1}|}\sqrt{1+|\lambda_{2}|}}
\right) \hspace*{0.7cm} \nonumber
 \\ =
 \delta_{m_{1},m_{2}} \langle {\bf R}_{1}| {\bf R}_{2} \rangle \delta_{\lambda_{1},\lambda_{2}}. \hspace*{3.5cm}
\end{eqnarray}

For convenience, in the next section we shall condense the full
set of quantum numbers $|m,{\bf R},\lambda \rangle$ into the single notation
$|\nu \rangle$. Therefore the sum over quantum numbers $\nu$ will stand for
\begin{eqnarray}
\sum_{\nu}=\int \!\!\! \frac{d^{2} {\bf R}}{2 \pi l_{B}^{2}}
\sum_{m=0}^{+\infty} \sum_{\lambda}.
\end{eqnarray}
It is finally straightforward to prove that the set of graphene vortex
states $|m,{\bf R},\lambda \rangle$ obeys a completeness relation, which reads
\begin{eqnarray}
\lefteqn{
\sum_\nu |\nu \rangle \langle \nu| =
 \int \!\!\! \frac{d^{2} {\bf R}}{2 \pi l_{B}^{2}}
\sum_{m=0}^{+ \infty}
\sum_{\lambda} \frac{1}{1+|\lambda|} }
\nonumber
\\ & \times
\left(
\begin{array}{cc}
\lambda^{2} |m-1,{\bf R} \rangle \langle m-1,{\bf R} | & - i \lambda |m-1,{\bf R} \rangle \langle m,{\bf R} | \\
i \lambda |m,{\bf R} \rangle \langle m-1,{\bf R} | & |m,{\bf R} \rangle \langle m,{\bf R} |
\end{array}
\right) \nonumber \\
& = \displaystyle{
\int \!\!\! \frac{d^{2} {\bf R}}{2 \pi l_{B}^{2}}} \sum_{m=0}^{+ \infty}
\left(
\begin{array}{cc}
|m,{\bf R} \rangle \langle m,{\bf R} | & 0 \\
0 & |m,{\bf R} \rangle \langle m,{\bf R} |
\end{array}
\right)
 \nonumber
\\
& =
\left(
\begin{array}{cc}
1 & 0 \\
0 & 1
\end{array}
\right), \hspace*{5.3cm}
 \label{comp}
\end{eqnarray}
where we have used the completeness relation (\ref{compvortex}) satisfied by the
vortex states.

\section{General formalism for a smooth potential}
\label{sec:formalism}

\subsection{Matrix elements of the potential}
\label{Matrix}

In order to investigate the effect of a smooth potential under magnetic field,
we shall naturally project the different contributions of Hamiltonian
(\ref{prob}) in the graphene vortex representation. Although being basic, this
projection sheds already interesting light on the different processes at play
and shows the essential differences between the different kinds of potentials
that may be encountered in graphene, see Eq. (\ref{prob2}). Using Eq.
(\ref{Dirac}), the matrix elements of the diagonal part of the potential (i.e.,
associated to scalar and mass potentials) can be written as
\begin{eqnarray}
\lefteqn{\langle \nu_{1} | V_{\mathrm{diag}} |\nu_{2} \rangle
= \left[
(1+|\lambda_{1})|(1+|\lambda_{2}|)
\right]^{-1/2}
\nonumber} \hspace*{1cm}
\\&
\times \left\{
\lambda_{1} \lambda_{2} \langle m_{1}-1,{\bf R}_{1}|V_{s}+V_{z}|m_{2}-1,{\bf R}_{2}
\rangle
\right. \nonumber \hspace*{0.5cm} \\
&
\left.
+ \langle m_{1},{\bf R}_{1}|V_{s}-V_{z}|m_{2},{\bf R}_{2}
\rangle
\right\}
. \hspace*{1.3cm}
\label{mat}
\end{eqnarray}
The off-diagonal terms of the potential (i.e., the random vector potential contribution) give rise
to the following matrix elements:
\begin{eqnarray}
\lefteqn{\langle \nu_{1}| V_{\mathrm{off}} |\nu_{2} \rangle
=
i \left[
(1+|\lambda_{1}|)(1+|\lambda_{2})
\right]^{-1/2}} \hspace*{1cm}
\nonumber \\
& \times
\left\{
 \lambda_{1} \langle m_{1}-1,{\bf R}_{1}|V_{x}-iV_{y}|m_{2},{\bf R}_{2}
\rangle \nonumber \hspace*{1.3cm}
\right.
\\
&
\left.
- \lambda_{2} \langle m_{1},{\bf R}_{1}|V_{x}+ iV_{y}|m_{2}-1,{\bf R}_{2}
\rangle
\right\}
.
\label{matoff} \hspace*{0.4cm}
\end{eqnarray}

We have shown in Ref. \onlinecite{Champel2007} that it is possible to evaluate
exactly the matrix elements of a smooth function $V({\bf r})$ in the vortex
representation [provided that $V(x,y)$ is an analytic function of both $x$ and
$y$] and write them as a series in powers of the magnetic length $l_{B}$,
\begin{eqnarray}
\langle m_{1},{\bf R}_{1}|V|m_{2},{\bf R}_{2}
\rangle
=\langle {\bf R}_{1} | {\bf R}_{2}\rangle \,
v_{m_{1};m_{2}}\left( {\bf R}_{12}\right)
\label{smallv}
\end{eqnarray}
with ${\bf R}_{12}=\left[
{\bf R}_{1}+{\bf R}_{2}+i ({\bf R}_{2}-{\bf R}_{1}) \times \hat{{\bf z}}
\right]/2$ and
\begin{eqnarray}
\label{general}
v_{m_{1};m_{2}}({\bf R}) &=& \int d^2 {\bm \eta} \,
\Psi_{m_1,{\bf R}}^{\ast}({\bm \eta}) \Psi_{m_2,{\bf R}}({\bm \eta})
V({\bm \eta}) \\
&=&
\sum_{j=0}^{+ \infty}
\left(\frac{l_{B}}{\sqrt{2}} \right)^{j} v^{(j)}_{m_{1},m_{2}}({\bf R}),
\label{series}
\\
v^{(j)}_{m_{1};m_{2}}({\bf R})
&=&
\sum_{k=0}^{j}
\frac{(m_{1}+k)!}{\sqrt{m_{1}!m_{2}!}}
\frac{\delta_{m_{1}+k,m_{2}+j-k}}{k!(j-k)!}
\label{space} \\
&& \times
\left( \partial_{X}+i\partial_{Y}\right)^{k}
\left( \partial_{X}-i\partial_{Y}\right)^{j-k}
V({\bf R}). \nonumber
\end{eqnarray}
Clearly, the use of an analytical expansion around the complex point ${\bf R}_{12}$ in Eq. (\ref{smallv}) puts some constraints on the types of potential that can be considered in the present formalism. We emphasize that relation (\ref{smallv}) holds for any {\em physical} potentials $V$ (which are necessarily smooth functions of the space variables).
In contrast, pointlike (i.e., zero-range) potentials involving Dirac delta functions which represent toy models simulating short-range potentials can not be treated within the present formalism.
If the magnetic length $l_{B}$ corresponds to the shortest length scale [here,
basically, $l_{B}$ has to be compared with the characteristic length scale of
spatial variations in the function $V({\bf R})$, see Eqs. (\ref{series}) and (\ref{space})],
we see that we have naturally ordered the different contributions to the matrix
elements by their order of magnitude in high magnetic fields.

At leading order ($l_{B} \to 0$), we get from Eqs. (\ref{mat})-(\ref{space}) for
coinciding vortex positions ${\bf R}_{1}={\bf R}_{2}={\bf R}$,
\begin{eqnarray}
\langle \nu_{1}| V_{\mathrm{diag}} |\nu_{2} \rangle
 \approx
\delta_{m_{1},m_{2}} \left[ \delta_{\lambda_{1},\lambda_{2}}
\, V_{s}\left( {\bf R} \right)
-\delta_{\lambda_{1},-\lambda_{2}} \, V_{z}\left( {\bf R} \right)
\right]
. \nonumber \\
\label{Vdiag0}
\end{eqnarray}
We remark that in the limit $l_{B} \to 0$ the diagonal elements $V_{s}$ and
$V_{z}$ of $V$ do not introduce a mixing between Landau levels. For smooth
functions $V_{x}$ and $V_{y}$ we get in the same limit $l_{B} \to 0$,

\begin{eqnarray}
 \lefteqn{ \langle \nu_{1} | V_{\mathrm{off}} |\nu_{2} \rangle
\approx - i \left[
(1+|\lambda_{1}|)
(1+|\lambda_{2}|)
 \right]^{-1/2} \nonumber } \hspace*{1cm} \\
&
\times
\left\{
\lambda_{2}\delta_{m_{1},m_{2}-1}\left[V_{x}({\bf R})+iV_{y}({\bf R}) \right]
\nonumber \right. \hspace*{2cm} \\
& \hspace*{0.1cm}
\left.
-\lambda_{1} \delta_{m_{1}-1,m_{2}}\left[V_{x}({\bf R})-iV_{y}({\bf R}) \right]
\right\}. \hspace*{1cm}
\label{Voff0}
\end{eqnarray}
We note with Eq. (\ref{Voff0}) that the off-diagonal elements $V_{x}$ and
$V_{y}$ do mix adjacent Landau levels already at leading order in $l_{B}$, in
contrast to the diagonal elements $V_{s}$ and $V_{z}$ of $V$. This difference
clearly calls for a different treatment of the diagonal and off-diagonal parts
of the total potential $V$. Off-diagonal contributions can be
treated perturbatively at high magnetic field by assuming that $V_{x}$ and
$V_{y}$ are small in amplitude in addition of being smooth functions at the
scale $l_{B}$. Such a constraint on the amplitude can be relaxed in the
treatment of the diagonal contributions of $V$.

The next (sub-dominant) contributions of order $l_{B}$ to the matrix elements of
$V_{\mathrm{diag}}$ are proportional to
\begin{eqnarray}
\delta_{m_{1}+1,m_{2}}
 (\partial_{X}+i\partial_{Y})
\left\{
\sqrt{m_{2}} \left[ V_{s}({\bf R})-V_{z}({\bf R})\right]
\nonumber \right. \hspace*{0.5cm}
\\
\left.
+\lambda_{1}\lambda_{2} \sqrt{m_{1}}\left[ V_{s}({\bf R})+V_{z}({\bf R})\right]
\right\}
+ c.c. (1 \leftrightarrow 2)
,
\label{v1}
\end{eqnarray}
where the notation $ c.c. (1 \leftrightarrow 2)$ means taking the complex
conjugate and exchanging the indexes 1 and 2 of the former expression. This
contribution induces a mixing between both adjacent Landau levels and band
indices $\lambda$. Moreover, the mixture of positive- and negative-energy
components stems from both components $V_{s}$ and $V_{z}$ of the potential
energy. It is interesting to note that for a large Landau-level index, the
mixture arising purely from $V_{s}$ (i.e., taking the mass term $V_{z}=0$)
gets negligible when $\lambda_{1} \lambda_{2}=-1$. For instance, when $m_{1}$
and $m_{2} \gg 1$, we have
\begin{equation}
\sqrt{m_{1}+1}+\lambda_{1}\lambda_{2} \sqrt{m_{1}} \approx \sqrt{m_{1}}
\left( 1+\lambda_{1}\lambda_{2}
\right)
\end{equation}
for the component $\delta_{m_{1}+1,m_{2}}$ of the matrix elements, Eq. (\ref{v1}),
associated with $V_{s}$. On the other hand, the band mixing becomes significant
for $m_{1}$ and $m_{2}$ close to 0. Specific signatures resulting from this
interband mixing, such as Zitterbewegung (or trembling motion) in a magnetic
field, have been discussed in the literature.
\cite{Rusin2008,Schliemann2008,Dora2009} By looking at next-order contributions
in $l_{B}^{2}$ for the matrix elements, we note that interband mixing occurs
also with the second derivatives of a pure scalar potential $V_{s}$ without
mixing the Landau levels. These mixing processes will be analyzed further 
in Sec. \ref{sec:high}.

\subsection{Green's-function formalism}

The nonorthogonality of the graphene vortex states preventing us to build a
wave-function perturbation theory, we shall instead use a Green's-function
formalism to get a more quantitative insight on the effect of a smooth
potential, following Refs. \onlinecite{Champel2008} and
\onlinecite{Champel2009}. Although the derivation of the equations of motion for
the graphene Green's function is very similar to that for the 2DEG Green's
function, we shall nevertheless describe the principal steps with some detail
here, in order to make this paper self-contained (we shall, however, not
reproduce the very technical details).

Retarded and advanced Green's functions are, respectively, defined as
\begin{eqnarray}
G^{R}(x_{1},x_{2}) &=& - i \theta(t_{1}-t_{2}) \langle
\left\{ \mathcal{\psi}(x_{1}), \mathcal{\psi}^{\dagger}(x_{2})\right\} \rangle, \\
G^{A}(x_{1},x_{2}) &=& i \theta(t_{2}-t_{1}) \langle
\left\{ \mathcal{\psi}(x_{1}), \mathcal{\psi}^{\dagger}(x_{2})\right\} \rangle,
\end{eqnarray}
where $\{,\}$ means the anti-commutator, and $\theta$ the Heaviside step
function [i.e., $\theta(t)=0$ for $t<0$ and $\theta(t)=1$ for $t>0$]. The
averages are evaluated in the grand canonical ensemble. The Green's functions
relate the field operator $\mathcal{\psi}(x_{1})$ of the particle at one point
$x_{1}=({\bf r}_{1},t_{1})$ in space-time to the conjugate field operator
$\mathcal{\psi}^{\dagger}(x_{2})$ at another point $x_{2}=({\bf r}_{2},t_{2})$.
The field operators $\mathcal{\psi}(x_{1})$ and
$\mathcal{\psi}^{\dagger}(x_{2})$ are expressed in terms of the eigenfunctions
$\tilde{\Psi}_{\nu}({\bf r})$ and eigenvalues $E_{\nu}$ as
\begin{eqnarray}
\mathcal{\psi}(x_{1}) &=& \sum_{\nu} c_{\nu} \tilde{\Psi}_{\nu}({\bf r}_{1}) \, e^{-iE_{\nu}t_{1}/\hbar},
\\
\mathcal{\psi}^{\dagger}(x_{2}) &=& \sum_{\nu} c^{\dagger}_{\nu} \tilde{\Psi}^{\dagger}_{\nu}({\bf r}_{2}) \, e^{iE_{\nu}t_{2}/\hbar},
\end{eqnarray}
where $c^{\dagger}_{\nu}$ and $c_{\nu}$ are, respectively, the creation and destruction operators.

As a basis of states, we shall then use the graphene vortex states $|\nu \rangle
=|m,{\bf R},\lambda \rangle$ which are eigenstates of Hamiltonian $H_{0}$ [Eq.
(\ref{Hgraphene})]. It is worth noting that, although these states $|\nu
\rangle$ are nonorthogonal, the associated creation and destruction operators
$c^{\dagger}_{\nu}$ and $c_{\nu}$ obey the usual algebra with the
anti-commutation rules
$\{c^{\dagger}_{\nu_{1}},c^{\dagger}_{\nu_{2}}\}=\{c_{\nu_{1}},c_{\nu_{2}}\}=0$
and $\{c_{\nu_{1}},c^{\dagger}_{\nu_{2}}\}=\delta_{\nu_{1},\nu_{2}}$.

Completeness relation (\ref{comp}) allows us to express the Green's function in
the graphene vortex representation, which we note
$G^{R,A}(\nu_{1},t_{1};\nu_{2},t_{2})=G^{R,A}(m_{1},{\bf
R}_{1},\lambda_{1},t_{1} ;m_{2},{\bf R}_{2},\lambda_{2},t_{2})$. Transposing its
definition originally made in terms of the electronic coordinates $({\bf r},t)$
into the vortex language, the latter Green's function gives the probability
amplitude for a vortex with circulation $m_{1}$ and band index $\lambda_{1}$
that is initially at position ${\bf R}_{1}$ at time $t_{1}$ to be at point ${\bf
R}_{2}$ at time $t_{2}$ with a new circulation $m_{2}$ and a band index
$\lambda_{2}$.
After Fourier transformation with respect to the time difference
$t=t_{1}-t_{2}$, the Green's function (denoted by $G_{0}$) corresponding to
Hamiltonian $H_{0}$ [i.e., Hamiltonian (\ref{prob}) with $V=0$] are written in
the energy ($\omega$) representation as
\begin{eqnarray}
G_{0}^{R,A}(\nu_{1};\nu_{2})=\frac{\delta_{m_{1},m_{2}} \,
\delta_{\lambda_{1},\lambda_{2}} \,\langle {\bf R}_{1} | {\bf R}_{2}
\rangle}{\omega-E_{m_{1},\lambda_{1}} \pm i 0^{+}}.
\end{eqnarray}

Retarded and advanced Green's function in the presence of the smooth potential
$V$ are obtained from Dyson equation, which takes the following form in the
$\nu$ representation (we again considered the Fourier transform of Green's
function with respect to time difference)
\begin{eqnarray}
\left(
\omega - E_{m_{1},\lambda_{1}} \pm i 0^{+}
\right) G^{R,A}(\nu_{1};\nu_{2})
=\langle \nu_{1}| \nu_{2} \rangle \hspace*{1cm} \nonumber \\
+\sum_{\nu_{3}} V_{\nu_{1};\nu_{3}} G^{R,A}(\nu_{3};\nu_{2})
. \label{Dyson}
\end{eqnarray}
Here the general matrix elements $V_{\nu_{1};\nu_{2}}=\langle \nu_{1} |V|\nu_{2}
\rangle= \langle {\bf R}_{1} | {\bf R}_{2} \rangle \,
v_{m_{1},\lambda_{1};m_{2},\lambda_{2}}({\bf R}_{12}) $ are given by expressions
(\ref{mat})-(\ref{space}). For $V\neq 0$, the graphene vortex Green's function
is generally no more diagonal with respect to the quantum numbers $m$ and
$\lambda$, and the mixing between the different quantum numbers depend on the
characteristic properties of the potential $V$. However, it turns out that, as a
result of the coherent states character with respect to vortex position ${\bf
R}$ encompassed within overlap (\ref{overlapR}), the propagation of the graphene
vortex Green's function with respect to vortex positions ${\bf R}_{1}$ and ${\bf
R}_{2}$ is constrained to necessarily take the form
\begin{eqnarray}
G(\nu_{1};\nu_{2})= \langle {\bf R}_{1} |{\bf R}_{2} \rangle \,
g_{m_{1},\lambda_{1} ;m_{2},\lambda_{2}}
\left(
{\bf R}_{12}
\right)
 ,
 \label{Gvortexnonlocal}
\end{eqnarray}
similarly to the matrix elements of the potential [see Eq. (\ref{smallv})].
Such exact dependence, Eq. (\ref{Gvortexnonlocal}), can be derived from Dyson Eq.
(\ref{Dyson}) in the same way as done in Ref. \onlinecite{Champel2008}.
Remarkably, it implies that the nonlocal graphene Green's function
$G(\nu_{1};\nu_{2})$ will be entirely determined once it is known at coinciding
vortex positions ${\bf R}_{1}={\bf R}_{2} \equiv {\bf R}$, and this result holds
irrespective of the potential $V$. It is then sufficient to consider Eq.
(\ref{Dyson}) for coinciding vortex positions. Because the derivation is the
same as for the 2DEG, we briefly outline here the last step leading to the final
equation of motion governing the function
$g_{m_{1},\lambda_{1};m_{2},\lambda_{2}}({\bf R})$ and refer the reader to
Sec. II of Ref. \onlinecite{Champel2008} for the mathematical details. The
nonlocal dependencies of the functions $G(\nu_{3};\nu_{2})$ and
$V_{\nu_{1};\nu_{3}}$ on the vortex positions which are known according to
relations (\ref{Gvortexnonlocal}) and (\ref{smallv}) are exploited to evaluate
the integral over the continuous variable ${\bf R}_{3}$ on the right-hand side
of Eq. (\ref{Dyson}). This integral then transforms into a series expansion in
powers of $l_{B}$.
We obtain that Dyson equation for the retarded graphene vortex Green's function $g({\bf R})$
(from now on, we drop the $R$ upperscript associated to retarded) corresponding to 
Hamiltonian (\ref{prob}) reads
\begin{eqnarray}
(\omega-E_{m_{1},\lambda_{1}}+ i 0^{+})
g_{m_{1},\lambda_{1};m_{2},\lambda_{2}}({\bf R})
= \delta_{m_{1},m_{2}} \, \delta_{\lambda_{1},\lambda_{2}}
\nonumber
\\
+
\sum_{k=0}^{+\infty}\left(\frac{l_{B}}{\sqrt{2}} \right)^{2k} \frac{1}{k!}
\sum_{m_{3},\lambda_{3}}
\left(\partial_{X}-i\partial_{Y } \right)^{k}
v_{m_{1},\lambda_{1};m_{3},\lambda_{3}}({\bf R})
\nonumber
\\
\times
\left(\partial_{X}+i\partial_{Y } \right)^{k}
g_{m_{3},\lambda_{2};m_{2},\lambda_{2}} ({\bf R}) \hspace*{0.5cm}
\label{Dyson4}
\end{eqnarray}
with $E_{m,\lambda}=\lambda E_{m}=\lambda \hbar \sqrt{2m} v_{F}/l_{B}$.

Another important aspect of the change in function (\ref{Gvortexnonlocal}), which
appears clearly with the form (\ref{Dyson4}) of Dyson equation and with
expressions (\ref{series}) and (\ref{space}) for the matrix elements of the
potential taken at coinciding vortex positions, is that the nonanalytic
dependence of the nonlocal graphene vortex Green's function $G(\nu_{1};\nu_{2})$
on the magnetic $l_{B}$ has been entirely extracted [in formula
(\ref{Gvortexnonlocal}), this nonanalytic dependence is only contained in the
overlap $\langle {\bf R}_{1} | {\bf R}_{2} \rangle$]. In other terms, the
function $g({\bf R})$ is obviously analytic in $l_{B}$ and thus well behaves in
the semiclassical limit of zero magnetic length ($l_{B} \to 0$). This property
can be used to solve Eq. (\ref{Dyson4}) order by order in powers of $l_{B}$ and
thus to provide a semiclassical expansion of the graphene vortex Green's
function $g_{m_{1},\lambda_{1};m_{2},\lambda_{2}}$ as
\begin{equation}
g_{m_{1},\lambda_{1};m_{2},\lambda_{2}}=\sum_{j=0}^{+\infty}
\left(\frac{l_{B}}{\sqrt{2}} \right)^{j} g^{(j)}_{m_{1},\lambda_{1}; m_{2},\lambda_{2}}.
\label{expg}
\end{equation}
 Because the series, Eq. (\ref{expg}), is then only asymptotic in nature (the obtained
solution holds in the limit $l_{B} \to 0$, but is not controlled at finite
$l_{B}$), we aim here at solving directly and non-perturbatively in $l_B$
Dyson Eq. (\ref{Dyson4}).

For this purpose, we have found in Ref. \onlinecite{Champel2009} that it is very
convenient to introduce the simultaneous changes in functions,
\begin{eqnarray}
\tilde{g}_{m_{1},\lambda_{1};m_{2},\lambda_{2}}({\bf R}) &=& e^{-(l_{B}^{2}/4) \Delta_{{\bf R}}} g_{m_{1},\lambda_{1};m_{2},\lambda_{2}}({\bf R}), \label{changeg} \\
\tilde{v}_{m_{1},\lambda_{1};m_{2},\lambda_{2}}({\bf R}) &=& e^{-(l_{B}^{2}/4) \Delta_{{\bf R}}} v_{m_{1},\lambda_{1};m_{2},\lambda_{2}}({\bf R}), \label{changev}
\end{eqnarray}
where the symbol $\Delta_{{\bf R}}$ means the Laplacian operator taken with
respect to the vortex position ${\bf R}$. After substitution of these
expressions (\ref{changeg}) and (\ref{changev}) into Eq. (\ref{Dyson4}), we get a
new equation for the unknown function
$\tilde{g}_{m_{1},\lambda_{1};m_{2},\lambda_{2}}({\bf R})$ with a higher-order 
differential operator than the one appearing on the right-hand side of Eq.
(\ref{Dyson4})
\begin{eqnarray}
(\omega-E_{m_{1},\lambda_{1}}+ i 0^{+})
\tilde{g}_{m_{1},\lambda_{1};m_{2},\lambda_{2}}({\bf R})
=
 \delta_{\lambda_{1},\lambda_{2}} \delta_{m_{1},m_{2}} \nonumber \\
+
\sum_{m_{3},\lambda_{3}}
\tilde{v}_{m_{1},\lambda_{1};m_{3},\lambda_{3}}({\bf R})
\star
\tilde{g}_{m_{3},\lambda_{3};m_{2},\lambda_{2}} ({\bf R}), \hspace*{0.5cm}
\label{Dysonfinal}
\end{eqnarray}
where the symbol $\star$ stands for the bi-differential operator defined by
\begin{eqnarray}
\star=\exp\left[i \frac{l_{B}^{2}}{2} \left( \overleftarrow{\partial}_{X} \overrightarrow{ \partial}_{Y}
- \overleftarrow{\partial}_{Y} \overrightarrow{ \partial}_{X} \right) \right].
\label{star}
\end{eqnarray}
The arrow above the partial derivatives indicates to which side the derivative
acts. Note that the passage from Eq. (\ref{Dyson4}) to Eq. (\ref{Dysonfinal})
is more straightforward by going to Fourier space (see Appendix A of Ref.
\onlinecite{Champel2009}). It is worth mentioning that, by starting from the
other Dyson equation (i.e., formally $G=G_{0}+GVG_{0}$) and following the same
steps as detailed previously, we can derive a second equation satisfied by the
function $\tilde{g}$,
\begin{eqnarray}
(\omega-E_{m_{2},\lambda_{2}}+ i 0^{+})
\tilde{g}_{m_{1},\lambda_{1};m_{2},\lambda_{2}}({\bf R})
=
 \delta_{\lambda_{1},\lambda_{2}} \delta_{m_{1},m_{2}} \nonumber \\
+
\sum_{m_{3},\lambda_{3}}
\tilde{g}_{m_{1},\lambda_{1};m_{3},\lambda_{3}}({\bf R})
\star
\tilde{v}_{m_{3},\lambda_{3};m_{2},\lambda_{2}} ({\bf R}). \hspace*{0.5cm}
\label{Dysonfinalbis}
\end{eqnarray}
The particular form \cite{noteRaikh} of exact Eqs.
(\ref{Dysonfinal})-(\ref{Dysonfinalbis}), reminiscent of the so-called
star-product, will be further used and commented in Secs. \ref{sec:high} and \ref{sec:deformation}.

In order to compute local physical observables such as the local density of
states, we need to express Green's function in terms of the electronic
positions ${\bf r}$. The electronic Green's function is a 2 x 2 matrix in the
pseudo-spin space, and is defined as $\hat{G}({\bf r},{\bf r}')=\langle {\bf r}
|\hat{G} |{\bf r}'\rangle$. At a practical level, it is useful to directly
relate the nonlocal electronic Green's function to the {\it local} graphene vortex 
Green's function $g_{m_{1},\lambda_{1};m_{2},\lambda_{2}}({\bf R})$ (at coinciding vortex
positions) or alternatively to the modified vortex Green's function
$\tilde{g}_{m_{1},\lambda_{1};m_{2},\lambda_{2}}({\bf R})$. First, the
electronic Green's function can be straightforwardly linked to the nonlocal
graphene vortex Green's function $G(\nu_{1};\nu_{2})$ through a change in
representation which is performed by using twice completeness relation
(\ref{comp}). Then, using Eq. (\ref{Gvortexnonlocal}) and following the
calculations made in Ref. \onlinecite{Champel2008} for the 2DEG, we get the
following relation
\begin{widetext}
\begin{eqnarray}
\hat{G}({\bf r},{\bf r}',\omega)
= \int \!\!\!
\frac{d^{2}{\bf R}}{2 \pi l_{B}^{2}}
\sum_{m_{1},\lambda_{1}} \sum_{m_{2},\lambda_{2}}
\left(
\begin{array}{cc}
\lambda_{1}\lambda_{2}
\Psi^{\ast}_{m_{2}-1,{\bf R}}
({\bf r}')
\Psi_{m_{1}-1,{\bf R}}
({\bf r}) & - i \lambda_{1} \Psi^{\ast}_{m_{2},{\bf R}}
({\bf r}')
\Psi_{m_{1}-1,{\bf R}}
({\bf r})
\\
i \lambda_{2} \Psi^{\ast}_{m_{2}-1,{\bf R}}
({\bf r}')
\Psi_{m_{1},{\bf R}}
({\bf r})
&
\Psi^{\ast}_{m_{2},{\bf R}}
({\bf r}')
\Psi_{m_{1},{\bf R}}
({\bf r})
\end{array}
\right)
\nonumber \\
\times
e^{ -(l_{B}^{2}/2) \Delta_{{\bf R}}} \left[
\frac{g_{m_{1}, \lambda_{1};m_{2},\lambda_{2}}({\bf R})}
{
\sqrt{1+|\lambda_{1}|}
\sqrt{1+|\lambda_{2}|}
} \right]
,
\label{passage}
\end{eqnarray}
where the functions $\Psi_{m,{\bf R}}({\bf r})$ correspond to the so-called
vortex wave functions written in Eq. (\ref{Sol2}). Inverting expression
(\ref{changeg}), i.e., writing $g_{m_{1}, \lambda_{1};m_{2},\lambda_{2}}({\bf
R}) = e^{(l_{B}^{2}/4) \Delta_{\bf R} } \,\tilde{g}_{m_{1},
\lambda_{1};m_{2},\lambda_{2}}({\bf R})$ and inserting this result into Eq.
(\ref{passage}), we get after integrations by parts (so that the operator
involving the Laplacian acts on the product of wave functions rather on the
local vortex Green's function)
\begin{eqnarray}
\hat{G}({\bf r},{\bf r}',\omega)
= \int \!\!\!
\frac{d^{2}{\bf R}}{2 \pi l_{B}^{2}}
\sum_{m_{1},\lambda_{1}} \sum_{m_{2},\lambda_{2}}
e^{ - (l_{B}^{2}/4) \Delta_{{\bf R}}}
\left(
\begin{array}{cc}
\lambda_{1}\lambda_{2}
\Psi^{\ast}_{m_{2}-1,{\bf R}}
({\bf r}')
\Psi_{m_{1}-1,{\bf R}}
({\bf r}) & - i \lambda_{1} \Psi^{\ast}_{m_{2},{\bf R}}
({\bf r}')
\Psi_{m_{1}-1,{\bf R}}
({\bf r})
\\
i \lambda_{2} \Psi^{\ast}_{m_{2}-1,{\bf R}}
({\bf r}')
\Psi_{m_{1},{\bf R}}
({\bf r})
&
\Psi^{\ast}_{m_{2},{\bf R}}
({\bf r}')
\Psi_{m_{1},{\bf R}}
({\bf r})
\end{array}
\right)
\nonumber \\
\times
\frac{\tilde{g}_{m_{1}, \lambda_{1};m_{2},\lambda_{2}}({\bf R})}
{
\sqrt{1+|\lambda_{1}|}
\sqrt{1+|\lambda_{2}|}
}
.
\label{passage2}
\end{eqnarray}
Because the functions $\tilde{g}_{m_{1},\lambda_{1};m_{2},\lambda_{2}}$ may
depend on $\lambda_{1}$ and $\lambda_{2}$, the electronic Green's function
$\hat{G}({\bf r},{\bf r}')$ possesses, in general, off-diagonal elements.
The above equation is a central one, because it shows that any physical
observable can be computed from the knowledge of the {\it local} vortex
Green's function $\tilde{g}_{m_{1}, \lambda_{1};m_{2},\lambda_{2}}({\bf R})$.
\end{widetext}

\section{High magnetic field regime \label{high}}
\label{sec:high}
\subsection{Regime of negligible Landau-level mixing}

While Eqs. (\ref{Dysonfinal})-(\ref{Dysonfinalbis}) can, in principle, be
considered for any magnetic fields, we shall investigate here the regime of high
magnetic field only, for which Landau level mixing can be safely neglected. This
regime can be reached under reasonable conditions (i.e., for fields on the order
of 1 T or higher) provided that the potential landscape is sufficiently smooth.
Indeed, Landau-level mixing processes are described within Eq. (\ref{Dysonfinal}) by the
matrix elements $\tilde{v}_{m_{1},\lambda_{1};m_{3},\lambda_{3}}$ with $m_{1}
\neq m_{3}$. From the expressions of the matrix elements of the potential
coupling adjacent Landau levels calculated in the vortex representation in Sec.
\ref{Matrix}, we can formulate a clear quantitative criterion for neglecting
Landau-level mixing due to the diagonal contributions of the potential $V$ in
graphene,
\begin{equation}
l_{B} \left|{\bm \nabla}_{{\bf R}}V({\bf R})\right| \ll
\left(\sqrt{m+1}-\sqrt{m} \right)\hbar \Omega_{c}.
\label{ineq1}
\end{equation}
In graphene and for a field of 5 T, we have
$\hbar \Omega_{c}=\hbar \sqrt{2}v_{F}/l_{B} \approx 80$ meV
and $l_B \simeq 11$ nm. Recent experimental STS measurements of the
spatial dispersion of Landau levels in epitaxial graphene \cite{Miller2009} give
at most typical linear variations in $\delta V \simeq 5$ meV on length scales
$\xi \simeq 20$ nm.
Thus $l_B |{\bm \nabla} V|/\hbar \Omega_{c} \lesssim l_B \delta V/ \xi \hbar
\Omega_c \simeq 0.03$, a very small number indeed, so that the 
limit of negligible Landau level mixing is well obeyed.
We shall furthermore suppose that the Landau level mixing processes due to the
off-diagonal part of $V$ are small. According to Eq. (\ref{Voff0}), this implies
\begin{eqnarray}
\left| V_{x,y}({\bf R})\right| \ll \left(\sqrt{m+1}-\sqrt{m} \right) \hbar \Omega_{c} .
\label{ineq2}
\end{eqnarray}
Under inequalities (\ref{ineq1}) and (\ref{ineq2}), Landau-level mixing
processes due to the spatial variations in the scalar potential $V_{s}$ and of
the random mass $V_{z}$ or to the spatial fluctuations ($V_{x}$ and $V_{y}$) of
the vector potential are small and can be accounted for perturbatively on the
basis of Eq. (\ref{Dysonfinal}).

Henceforth, we shall concentrate on the main relevant processes occurring at
high magnetic field in a smooth potential. In this regime, the Landau-level
degeneracy is principally lifted by the presence of both the potentials $V_{s}$
and $V_{z}$, which give rise for $m \geq 1$ to the following diagonal
($m_1=m_2=m$) matrix elements in the vortex representation
\begin{eqnarray}
v_{m;\lambda_{1};\lambda_{2}}({\bf R})
&=
&
\delta_{m_{1},m_{2}} v_{m_{1},\lambda_{1};m_{2},\lambda_{2}}({\bf R})
\nonumber \\
&=&
\delta_{\lambda_{1},\lambda_{2}} \, v_{m}^{+}({\bf R})+\delta_{\lambda_{1},-\lambda_{2}}
\, v_{m}^{-} ({\bf R})
, \hspace*{0.5cm} \label{matdia}
\end{eqnarray}
where the diagonal and off-diagonal components of the potential matrix elements
in pseudospin space, respectively, read
\begin{eqnarray}
\nonumber
 v_{m}^{\pm}({\bf R}) &=&
\frac{1}{2} \int d^2 {\bm \eta} \left[ \,
|\Psi_{m,{\bf R}}({\bm \eta})|^2 
(V_s({\bm \eta})-V_z({\bm \eta})) \right. \\
\label{vplus}
&& \left. \pm 
|\Psi_{m-1,{\bf R}}({\bm \eta})|^2
(V_s({\bm \eta})+V_z({\bm \eta})) \right] \\
&=& \frac{1}{2}
\sum_{j=0}^{+\infty}
 \frac{(m+j)!}{m! (j!)^{2}} \left(\frac{l_{B}^{2}}{2} \Delta_{{\bf R}} \right)^{j}
\Big\{ \frac{}{} V_{s}({\bf R})-V_{z}({\bf R}) \nonumber \\
&& \pm \frac{m}{m+j}\left( V_{s}({\bf R})+V_{z}({\bf R}) \right)
\Big\}.
\label{vplusexpand}
\end{eqnarray}
To write down expressions (\ref{matdia})-(\ref{vplusexpand}), we have used Eqs.
(\ref{mat}) and (\ref{general})-(\ref{space}). We notice that even a {\it scalar}
potential $V_{s}$ introduces a coupling between the bands $\lambda=\pm$ for $m \geq 1$ 
through its nonlocal differential contributions arising with $j>1$. For
instance, a quadratic scalar potential generically mixes the positive and
negative energy components, even in the absence of a mass term ($V_{z}=0$). The
case $m=0$ has to be treated as a special case since there is only one band
($\lambda=0$ necessarily). The matrix elements for the lowest Landau level
$m=0$ read
\begin{eqnarray}
v_{0}({\bf R}) &=& 
\int d^2 {\bm \eta} 
|\Psi_{0,{\bf R}}({\bm \eta})|^2 
\left[V_s({\bm \eta})-V_z({\bm \eta}) \right] \\
& = & \sum_{j=0}^{+\infty}
 \frac{1}{j!} \left(\frac{l_{B}^{2}}{2} \Delta_{{\bf R}} \right)^{j}
[V_{s}({\bf R})-V_{z}({\bf R})]
. \label{v0}
\end{eqnarray}

We have seen previously that Dyson Eq.~(\ref{Dysonfinal}) is greatly 
simplified when considering modified matrix elements 
$\tilde{v}_{m,\lambda_{1};m,\lambda_{2}}({\bf R}) = e^{-(l_{B}^{2}/4)
\Delta_{{\bf R}}} v_{m,\lambda_{1};m,\lambda_{2}}({\bf R})$, which
constitute the {\it effective potential} in Landau level $m$.
Using results given in the Appendix B of Ref. \onlinecite{Champel2009},
we get the action of the exponential differential operator onto the
product of two vortex functions with identical Landau level $m$ and 
positions ${\bf r}$:
\begin{eqnarray}
 K_{m}({\bf R}-{\bf r}) &\equiv&
e^{-(l_{B}^{2}/4) \Delta_{{\bf R}}} |\Psi_{m,{\bf R}}({\bf r})|^2
\\
&=& \frac{1}{\pi m! l_{B}^{2}}
 \frac{\partial^{m}}{\partial s^{m}}
\left. \frac{e^{-A_{s} ({\bf R}-{\bf r})^{2}/l_{B}^{2}}}{1+s} \right|_{s=0},
\label{Km}
\end{eqnarray}
with $A_{s}=(1-s)/(1+s)$.
Thus, the diagonal and off-diagonal effective potentials (in pseudospin space) 
read for $m\geq1$,
\begin{eqnarray}
\nonumber
\tilde{v}_{m}^{\pm}({\bf R}) &=&
\frac{1}{2} \int d^2 {\bm \eta} \left[ \,
K_m({\bf R}-{\bm \eta})
(V_s({\bm \eta})-V_z({\bm \eta})) \right. \\
&& \left. \pm 
K_{m-1}({{\bf R}-\bm \eta})
(V_s({\bm \eta})+V_z({\bm \eta})) \right].
\label{tildev}
\end{eqnarray}
We emphasize that formula~(\ref{tildev}) is non-perturbative 
in $l_B$ and possibly applies for potentials $V_s$ and $V_z$ with sizeable 
variations at the scale of $l_B$.
The effective potential in the lowest Landau level is also
readily obtained as:
\begin{eqnarray}
\tilde{v}_{0}({\bf R}) &=& 
\int d^2 {\bm \eta} 
K_0({\bf R}-{\bm \eta})
\left[ V_s({\bm \eta})-V_z({\bm \eta}) \right] . 
\label{tildev0}
\end{eqnarray}

Obviously, we find that the modified Green's function becomes also diagonal with 
respect to the Landau-level quantum number at large magnetic field (yet at finite magnetic field),
\begin{eqnarray}
\tilde{g}_{m_{1},\lambda_{1};m_{2},\lambda_{2}}({\bf R})= \delta_{m_{1},m_{2}}
\, \tilde{g}_{m_{1};\lambda_{1};\lambda_{2}}({\bf R})
,
\end{eqnarray}
and is determined for $m \geq 1$ by Dyson equation,
\begin{eqnarray}
\lefteqn{(\omega-E_{m,\lambda_{1}}+ i 0^{+})
\tilde{g}_{m;\lambda_{1};\lambda_{2}}({\bf R})
=
 \delta_{\lambda_{1},\lambda_{2}}
 \nonumber} \\
&&
+ \tilde{v}_{m}^{+}({\bf R}) \star \tilde{g}_{m;\lambda_{1};\lambda_{2}}({\bf R})
+
\tilde{v}_{m}^{-}({\bf R})
\star
\tilde{g}_{m;-\lambda_{1};\lambda_{2}} ({\bf R}) \hspace*{0.5cm}
\label{Dysondiag}
\end{eqnarray}
and for $m=0$ by

\begin{eqnarray}
(\omega+ i 0^{+})
\tilde{g}_{0}({\bf R})
&=&
1
+
\tilde{v}_{0}({\bf R})
\star
\tilde{g}_{0} ({\bf R}).
\label{Dyson0}
\end{eqnarray}
The other Dyson Eq. (\ref{Dysonfinalbis}) generates the different equation,
\begin{eqnarray}
\lefteqn{(\omega-E_{m,\lambda_{2}}+ i 0^{+})
\tilde{g}_{m;\lambda_{1};\lambda_{2}}({\bf R})
=
 \delta_{\lambda_{1},\lambda_{2}}}
 \nonumber \\&&
+ \tilde{g}_{m;\lambda_{1};\lambda_{2}}({\bf R}) \star \tilde{v}_{m}^{+} ({\bf R})
+
\tilde{g}_{m;\lambda_{1};-\lambda_{2}} ({\bf R})
\star \tilde{v}_{m}^{-}({\bf R}) \hspace*{0.5cm}
\label{Dysondiagbis}
\end{eqnarray}
for $m \geq 1$,
and

\begin{eqnarray}
(\omega+ i 0^{+})
\tilde{g}_{0}({\bf R})
&=&
1
+
\tilde{g}_{0}({\bf R})
\star
\tilde{v}_{0} ({\bf R})
\label{Dyson0bis}
\end{eqnarray}
for the lowest Landau level $m=0$.

\subsection{Locally flat potentials}
Now, we aim at solving Eqs. (\ref{Dysondiag})-(\ref{Dyson0bis}) at leading
order, which is vindicated when the potential is {\it locally} flat, i.e., when
potential curvature is small. This calculation includes the case of one-dimensional
potentials (i.e., {\it globally} flat potentials), for which the solution presented below
is exact. Indeed, as is clear from its explicit expression (\ref{star}), the
$\star$-bidifferential operator involves derivatives in two orthogonal positions.
In case where the potentials $V_{s}({\bf R})$ and $V_{z}({\bf R})$ are purely
one-dimensional potentials depending on the same coordinate, the function
$\tilde{g}({\bf R})$ will also only depend on the same and unique variable, so that the
$\star$ product between the functions $\tilde{v}$ and $\tilde{g}$ reduces to the
standard product of functions. In case of arbitrary spatial varying
two-dimensional potentials,
this constitutes a good approximation as long as temperature is higher than the
energy scales associated to local curvature terms, see Sec. \ref{hier} for
a general discussion. Dyson equation then is trivially solved, as the
system of differential Eqs. (\ref{Dysondiag})-(\ref{Dyson0bis})
transforms into a system of purely algebraic equations. Taking the difference
of Eqs. (\ref{Dysondiag}) and (\ref{Dysondiagbis}), we get for $m \geq 1$ the
relations between the different components of $\tilde{g}$,
\begin{eqnarray}
\tilde{g}_{m;+;-}({\bf R})&=&\tilde{g}_{m;-;+}({\bf R}) \label{R1} \\
\tilde{g}_{m;-;-}({\bf R})&=&\tilde{g}_{m;+;+}({\bf R})-2 \frac{E_{m;+}}
{\tilde{v}_{m}^{-}({\bf R})} \, \tilde{g}_{m;-;+} ({\bf R}) . \hspace*{0.5cm} \label{R2}
\end{eqnarray}
After simple algebra, we directly obtain the solution
\begin{eqnarray}
\!\tilde{g}_{m;\lambda_{1};\lambda_{2}}({\bf R}) &=&
\!\frac{1}{ \left[
\omega-\xi_{m,+}({\bf R})+i0^{+}
\right]
\left[
\omega-\xi_{m,-}({\bf R})+i0^{+}
\right]
}
\nonumber \\
&& \hspace{-2cm} \times \left\{ \left[\omega-\tilde{v}_{m}^{+}({\bf R})+E_{m,\lambda_{1}} \right]
\delta_{\lambda_{1},\lambda_{2}}
+\tilde{v}_{m}^{-}({\bf R}) \delta_{\lambda_{1},-\lambda_{2}} \right\}
\label{g1Dm}
\end{eqnarray}
with the poles (corresponding to the renormalized Landau levels) giving the
effective energies,
\begin{eqnarray}
\xi_{m,\pm}({\bf R})=
\tilde{v}_{m}^{+}({\bf R})\pm \sqrt{E_{m}^{2}+\left[\tilde{v}_{m}^{-}({\bf R}) \right]^{2}}.
\label{effective}
\end{eqnarray}
For $m=0$, the Green's function is characterized by a single pole and reads
\begin{eqnarray}
\tilde{g}_{0}({\bf R})&=& \frac{1}
{
\omega-\xi_{0}({\bf R})+i0^{+}
}
, \label{g1D0}
\end{eqnarray}
where $\xi_{0}({\bf R})=\tilde{v}_{0}({\bf R})$.

Equation~(\ref{effective}), with the explicit expression for the
renormalized potentials given in Eqs. (\ref{tildev}), provides
the leading result for the local Landau-level energy in arbitrary potentials
of diagonal type (i.e., scalar or mass-like). This
expression of course includes the case of a purely unidimensional (i.e., globally
flat) potential as an exact particular solution, but is a very good approximation
for smooth disordered potentials, which can be used to analyze
experimental STS results, as we discuss in Sec.~\ref{sec:local}.

\subsection{Locally curved potentials}
\label{curved}

This section presents the resolution of Dyson equation at next to leading
order, by extending the above calculation of the local vortex Green's function
to the incorporation of the effects of geometrical curvature in the potential
landscape. It has therefore a two-fold purpose. First, it provides a crucial
refinement of the previous expression~(\ref{g1Dm}), that includes important
quantum effects such as quantization of energy levels or tunneling
associated to the potentials $V_s$ and $V_z$, which are clearly missed in the
leading order guiding center Green's function.
Smaller energy scales associated to these physical processes are now accessible,
and the final expression will apply to arbitrary smooth potentials that are
{\it locally} curved. These important aspects are discussed in more detail in
Sec. \ref{hier}.
Second, in the special case of purely quadratic potentials (which thus have a {\it
global} constant curvature), the calculation provides essentially the exact Green's
function, from which one can gain interesting insights on the physics of confinement
or tunneling in graphene. We thus obtain analytically the quantization spectra of parabolic
quantum dots and show that the structure of energy levels qualitatively depends
on the type of confinement (electrostatic or mass type).
We henceforth assume that the diagonal potentials are {\it locally} well described
up to their second-order spatial derivatives.

\subsubsection{Lowest Landau level: Solution with both curved scalar
and mass potentials}

We start by considering the lowest Landau level $m=0$, which is the simplest
case to solve, as band indices are not involved. In that situation both
locally curved $V_{s}({\bf R})$ and $V_{z}({\bf R})$ can be solved altogether
(this is not the case for higher $m\geq1$ states, as will be discussed in
the next paragraphs).
Actually, Dyson Eq. (\ref{Dyson0}) for the lowest Landau level is formally
equivalent to the equations obtained \cite{Champel2009} for the 2DEG, as the
electrostatic potential $V({\bf R})$ for the 2DEG is just formally replaced by
the combination $V_{s}({\bf R})-V_{z}({\bf R})$ for graphene.
Working in the next to leading order, i.e., keeping local curvature terms of
order $l_B^4$ in the $\star$-bidifferential operator, Eq.~(\ref{star}), we can
directly transpose the solution of Ref. \onlinecite{Champel2009} to the graphene
case (for the method, see also Appendix~\ref{appA} of the present paper), which reads
\begin{eqnarray}
\tilde{g}_{0}({\bf R})
= - i \int_{0}^{+ \infty} \!\!\! \!\!\! dt \frac{
e^{i [\eta_0({\bf R})/\gamma_0({\bf R})] \left[t - \tau_0(t) \right]}
 }{\cos (\sqrt{\gamma_0({\bf R}) }t)}
e^{i t [\omega-\xi_{0}({\bf R})+i0^{+}]}
\nonumber
\\
\label{g0}
\end{eqnarray}
with
\begin{eqnarray}
\tau_0(t) &=& \frac{1}{\sqrt{\gamma_0({\bf R})}} \tan(\sqrt{\gamma_0({\bf R})}t) .
\label{tau}
\end{eqnarray}
The parameters $\gamma_0({\bf R})$ and $\eta_0({\bf R})$ in Eqs. (\ref{g0}) and (\ref{tau})
are geometric coefficients characterizing the local effective potential landscape
$\tilde{v}_0({\bf R})$ in the lowest Landau level:
\begin{eqnarray}
\gamma_0({\bf R}) &=& \frac{l_{B}^{4}}{4}
\left[
(\partial_{X}^{2}\tilde{v}_0)(\partial_{Y}^{2}\tilde{v}_0)-(\partial_{X}\partial_{Y}
\tilde{v}_0 )^{2}
\right]_{{\bf R}},
\label{gamma} \\
\eta_0({\bf R}) &=& \frac{l_{B}^{4}}{8}
\left[
(\partial_{X}^{2}\tilde{v}_0)(\partial_{Y}\tilde{v}_0)^{2}
+(\partial_{Y}^{2}\tilde{v}_0)(\partial_{X}\tilde{v}_0)^{2} \right. \nonumber \\
&& \left. \hspace*{0.7cm} -2(\partial_{X}\partial_{Y} \tilde{v}_0 )(\partial_{X}
\tilde{v}_0)(\partial_{Y} \tilde{v}_0) \right]_{{\bf R}}.
\label{eta}
\end{eqnarray}
The coefficient $\gamma_0({\bf R})$ is directly proportional to the Gaussian curvature of
the surface defined in the three-dimensional ``space'' $XYZ$ by the equation
$Z=\tilde{v}_0(X,Y)$. Its sign reflects the local topology of the effective
potential: $\gamma_0({\bf R}) >0$ indicates a locally elliptic potential with the presence of a local extremum (maximum or minimum), while $\gamma_0({\bf R}) <0$ corresponds to a locally hyperbolic (or saddle-shaped) potential. At the borders between the regions with curvatures with opposite signs, the potential is locally parabolic (the lines where the Gaussian curvature is zero are consequently called parabolic lines). For a complex disordered effective potential landscape, one expects that surface regions with positive and negative Gaussian curvature alternate.
Note that both cosine and tangent trigonometric functions in Eqs.
(\ref{g0}) and (\ref{tau}) transform into their hyperbolic counterparts in the
case $\gamma_0({\bf R}) <0$.
Equation (\ref{g0}) thus provides a general approximation scheme in the lowest
Landau level in the presence of arbitrary scalar and mass potentials that are
locally well described by local curvature coefficients, Eqs.~(\ref{gamma}) and (\ref{eta}).

Now, in the {\it particular} case of purely quadratic scalar and mass potentials, i.e.,
$V_s({\bf R})-V_z({\bf R})=V_{s0}-V_{z0}
+ \frac{1}{2} \left[({\bf R}-{\bf R}_{0}) \cdot {\bm \nabla}_{{\bf R}}
\right]^{2}(V_s-V_z) $, with ${\bf R}_{0}$ chosen as the single point where the
potential gradient vanishes, expression~(\ref{g0}) yields the {\it exact} Green's
function of the problem.
In that situation, the parameter $\gamma_0$ in Eq.~(\ref{gamma}) becomes
${\bf R}$-independent,
\begin{equation}
\gamma_0 = \frac{l_{B}^{4}}{4} \left\{
\partial_{X}^{2}(V_s-V_z)\partial_{Y}^{2}(V_s-V_z)
-\left[\partial_{X}\partial_Y (V_s-V_z)\right]^{2}
\right\},
\end{equation}
and describes the uniform (global) curvature of the potential, while the
${\bf R}$-independent part of the effective potential results from the simple
relation $ \tilde{v}_0 \equiv \tilde{v}_{0}({\bf R}_{0}) = 
\tilde{v}_0({\bf R})-\eta_0({\bf R})/\gamma_0=
V_{s0}-V_{z0} +(l_B^2/2)\Delta_{\bf R}(V_s-V_z)$.
For a confining potential, i.e., when $\gamma_0>0$, $\tau_0(t)$ is a $2\pi/\sqrt{\gamma_0}$
periodic function of time $t$. Direct Fourier analysis of expression~(\ref{g0})
using the above relations can be done and shows (see Appendix \ref{appC})
that the entire energy spectrum necessarily decomposes onto discrete modes:
\begin{eqnarray}
\label{spectrumg0}
E_{0,n} &=& \tilde{v}_0 + \mathrm{sgn}(\eta_0) \sqrt{\gamma_0} (2n+1)
\end{eqnarray}
with $n\geq0$ a positive integer, yielding a harmonic-oscillatorlike spectrum for the
parabolic quantum-dot model (in the large magnetic field regime considered here).
The general form of this spectrum will be discussed in the next section.
In contrast, for $\gamma_0<0$, the vortex Green's function expressed in the time
representation is no more periodic but decays on a time scale
$1/\sqrt{-\gamma_0}$, due to the cutoff function $1/\cosh(\sqrt{-\gamma_0}t)$.
These lifetime effects associated to negative Gaussian curvature are clear
manifestations of quantum tunneling in saddle-point potentials, and will be
considered in a future publication where transport properties in high magnetic field
will be considered.

\subsubsection{Arbitrary Landau level: Solution for a curved
scalar potential combined with a flat mass potential}

For $m \geq 1$, the structure of Dyson Eq. (\ref{Dysondiag}) for
graphene differs from that for the 2DEG case because of the possible
coupling between positive- and negative-energy bands. Two kinds of processes
are actually at work here.
First, non-zero mass potential $V_z$ directly couples the two bands, as
is clearly seen from the leading order Green's function in Eq.
(\ref{g1Dm}). Second and less obviously, higher-order
{\it scalar} processes can also induce band-mixing. Indeed, the effective
off-diagonal potential in Eq. (\ref{vplus}) reads in the small $l_B$-expansion:
$\tilde{v}_{m}^{-}({\bf R}) = -
V_{z}({\bf R})-m \frac{l_{B}^{2}}{2} \Delta_{{\bf R}}
V_{z}({\bf R})+\frac{l_{B}^{2}}{4} \Delta_{{\bf R}} V_{s}({\bf
R})+\mathcal{O}(l_B^4)$.
Thus, even for an identically zero mass term ($V_{z}=0$), positive-and
negative-energy bands are necessarily coupled by the second derivatives of
the scalar potential.

For reasons mentioned previously, one cannot analytically progress for
the $m\geq1$ Landau levels in case where both scalar and mass potentials
are strongly spatially dependent. In this section we therefore assume that
the scalar potential varies in space with sizeable local parabolic dispersion,
while the mass potential has much smoother spatial variations, so that local
derivatives of the mass term are associated to tiny energy scales (the reversed
situation, where the mass potential variations dominate the ones of the scalar potential,
is considered below in Sec.~\ref{parabolicmass}).
Since the calculation leading to the Green's function for graphene is largely
inspired from the 2DEG's derivation, \cite{Champel2009} details are produced
in Appendix \ref{appA}. The solution reads
\begin{eqnarray}
\nonumber
\tilde{g}_{m;\lambda_{1};\lambda_{2}}({\bf R})
&=&\!\! -\frac{i}{2}
\int_{0}^{+ \infty} \!\!\!\! \! dt
 \frac{
e^{i(\eta_m^+({\bf R})/\gamma_m^+({\bf R}))[t-\tau_m^+(t)]}}
{ \cos\Big(\sqrt{\gamma_m^+({\bf R})} t \Big)} \\
\nonumber
&& \times \sum_{\epsilon=\pm}
e^{i t (\omega -\xi_{m,\epsilon}({\bf R})+i0^{+})} \\
&& \hspace{-2cm} \times [ (1+\epsilon \lambda_{1} \alpha_{m}({\bf R}))
\delta_{\lambda_{1},\lambda_{2}}
+ \epsilon \beta_{m}({\bf R}) \delta_{-\lambda_{1},\lambda_{2}} ]
\label{solution}
\end{eqnarray}
where the effective energy $\xi_{m,\pm}({\bf R})$ is given by 
Eq. (\ref{effective}), and
\begin{eqnarray}
\alpha_{m}({\bf R}) &=& \frac{E_{m}}{\sqrt{E_{m}^{2}+\left[\tilde{v}_{m}^{-}({\bf R})
\right]^{2}}}, \label{alpha} \\
\beta_{m}({\bf R}) &=& \frac{\tilde{v}_{m}^{-}({\bf R})}{\sqrt{E_{m}^{2}+
\left[\tilde{v}_{m}^{-}({\bf R}) \right]^{2}}}.
\label{beta}
\end{eqnarray}
The geometric parameters $\gamma_m^+({\bf R})$ and $\eta_m^+({\bf R})$ have the same
definitions as in Eqs. (\ref{gamma}) and (\ref{eta}), where $\tilde{v}_0({\bf R})$ is
simply replaced by the effective potential $\tilde{v}_m^+({\bf R})$. The function $\tau_m^+(t)$ has also a similar
definition as in Eq.~(\ref{tau}) now in terms of $\gamma_m^+({\bf R})$.
Again, the above expression~(\ref{solution}) is quite general, and can be used
to describe arbitrary disordered (yet smooth) scalar potentials. A mass
contribution may be present, but only with negligible spatial variations for
the approximation to be valid.

Now, in the {\it particular} case where the bare scalar potential is globally quadratic
(i.e., has uniform curvature) and the mass potential is globally uniform, this
expression provides the {\it exact} Green's function. A possible parametrization
of such potentials reads $V_s({\bf R})=V_{s0}
+ \frac{1}{2} \left[({\bf R}-{\bf R}_{0}) \cdot {\bm \nabla}_{{\bf R}}
\right]^{2}V_s $ (with ${\bf R}_{0}$ chosen as the point where the
scalar potential gradient vanishes) and $V_z({\bf R}) = V_{z0}$.
The Gaussian curvature of the scalar potential becomes then constant and
independent of $m$,
\begin{equation}
\label{gammaplus}
\gamma^+ = \frac{l_{B}^{4}}{4} \left[
\partial_{X}^{2}V_s\partial_{Y}^{2}V_s
-(\partial_{X}\partial_Y V_s)^{2}
\right]
\end{equation}
while the ${\bf R}$-independent parts of the effective potentials read
$\tilde{v}_{m}^{+} \equiv \tilde{v}_{m}^{+}({\bf R}_{0})=V_{s0}+m (l_{B}^{2}/2) 
\Delta_{{\bf R}} V_{s}$
and $\tilde{v}_m^{-} = - V_{z0}+(l_{B}^{2}/4) \Delta_{{\bf R}} V_{s}$.
Fourier analysis as done in Appendix~\ref{appC} provides a spectrum of purely
discrete energy levels in the presence of 2D-parabolic scalar potential,
\begin{equation}
E_{m,n} = \tilde{v}_{m}^{+} \pm \sqrt{E_m^2+(\tilde{v}_m^-)^2} +
\mathrm{sgn}(\eta^+)\sqrt{\gamma^+} (2n+1).
\label{energy1}
\end{equation}
This form of quantization is quite reminiscent of the Fock-Darwin
spectrum for the non-relativistic 2DEG: besides the renormalization of Landau levels
(labeled by the integer $m$) due to the ${\bf R}$-independent part of the
effective potentials $\tilde{v}_m^+$ and $\tilde{v}_m^-$, the linear dependence
in the second discrete number $n$ provides an additional harmonic-oscillatorlike
contribution.
As a specific illustration for the case of a circular parabolic scalar
potential $V_s({\bf r}) = V_{s0} + (1/2) U_0 (x^2+y^2)$ together with a zero
mass term, one gets the following energy spectrum:
\begin{equation}
E_{m,n} = V_{s0}+ l_B^2 U_0(m+n+\frac{1}{2}) \pm 
\sqrt{(\hbar \Omega_c \sqrt{m})^2 +(l_B^2 U_0/2)^2},
\end{equation}
that we have already quoted in Eq.~(\ref{spectrum1}) in the large $\Omega_c$
limit.

\subsubsection{Arbitrary Landau level: Solution for a flat
scalar potential combined with a curved mass potential}
\label{parabolicmass}

We now consider the alternative solvable case of locally flat scalar potential,
together with a spatially dependent mass potential that can be locally well described
by a quadratic expansion. Solution of Dyson Eq.~(\ref{Dysondiag})
can then similarly be achieved, leading to the Green's function
for $m \geq 1$,
\begin{eqnarray}
\nonumber
\tilde{g}_{m;\lambda_{1};\lambda_{2}}({\bf R}) &=& -i
\int_{0}^{+ \infty} \!\!\! \! ds \frac{e^{i [\kappa_{m}({\bf R})+i0^+]s}}
{ \cos \left(\sqrt{\gamma_m^-({\bf R})}s\right)}
\Big\{ \cos[\theta_m(s)] \\
&&\hspace{-2.3cm}
\times \frac{\omega-\tilde{v}^{+}_{m}({\bf R})+\lambda_{1}E_{m}}{\kappa_m({\bf R})}
\delta_{\lambda_{1},\lambda_{2}} +i \sin[\theta_m(s)]\delta_{\lambda_{1},-\lambda_{2}} \Big\}
 \label{gVzquadra}
\end{eqnarray}
with $\theta_m(s) =  [\eta_m^-({\bf R})/\gamma_m^-({\bf R})][\tau_m^-(s)-s ]
- s \tilde{v}^{-}_{m}({\bf R})$ and
\begin{eqnarray}
\kappa_{m}({\bf R})=\mathrm{sgn}\left[\omega-\tilde{v}^{+}_{m}({\bf R})\right] \,
\left|\left(\omega-\tilde{v}^{+}_{m}({\bf R})\right)^{2}-E_{m}^{2} \right|^{1/2}
\label{kappa1}
\end{eqnarray}
if $|\omega-\tilde{v}^{+}_{m}({\bf R})| \geq E_{m}$, and
\begin{eqnarray}
\kappa_{m}({\bf R})=i\left|\left[\omega-\tilde{v}^{+}_{m}({\bf R})\right]^{2}-E_{m}^{2}
\right|^{1/2}
\label{kappa2}
\end{eqnarray}
if $|\omega-\tilde{v}^{+}_{m}({\bf R})| < E_{m}$.
Details for the derivation of result (\ref{gVzquadra}) can be found in Appendix
\ref{appB}. The geometric parameters $\gamma_m^-({\bf R})$ and $\eta_m^-({\bf R})$
in formula (\ref{gVzquadra}) are defined as in Eqs. (\ref{gamma}) and (\ref{eta})
with $\tilde{v}_0({\bf R})$ replaced by $-\tilde{v}_m^-({\bf R})$.
Taking the imaginary part of expression (\ref{gVzquadra}), we get that the local
density of states vanishes when $|\omega-\tilde{v}^{+}_{m}({\bf R})| < E_{m}$, meaning that
there are no states within this energy interval.

Now, in the {\it particular} case where the bare scalar potential is globally
uniform and the mass potential is globally quadratic, this
expression provides the {\it exact} Green's function in the absence of Landau-level mixing. A possible parametrization of such potentials reads
$V_s({\bf R})=V_{s0}$ and $V_z({\bf R}) = V_{z0} + \frac{1}{2} \left[({\bf R}-{\bf R}_{0}) \cdot
{\bm \nabla}_{{\bf R}} \right]^{2}V_z $ (with ${\bf R}_{0}$ chosen as the point where
the mass potential gradient vanishes).
The Gaussian curvature of the mass potential becomes then constant,
\begin{equation}
\gamma^- = \frac{l_{B}^{4}}{4} \left[
\partial_{X}^{2}V_z\partial_{Y}^{2}V_z
-(\partial_{X}\partial_Y V_z)^{2}
\right]
\end{equation}
while the ${\bf R}$-independent parts of the effective potentials read
$\tilde{v}_{m}^{+} = V_{s0}- (l_{B}^{2}/4) \Delta_{{\bf R}} V_{z}$
and $\tilde{v}_m^{-} \equiv \tilde{v}_m^{-}({\bf R}_{0})= - V_{z0}+m(l_{B}^{2}/2) \Delta_{{\bf R}} V_{z}$.
Fourier analysis as done before implies that the eigenenergies $\omega=E_{m,n}$
are determined by the implicit equation $\kappa_m-\tilde{v}_m^- =
\mathrm{sgn}(\eta^{-})\sqrt{\gamma^-}(2n+1) $ [we remind that the dependence on $\omega$ is contained
in $\kappa_m$, see Eq.~\ref{kappa1}], leading to the following discrete energy
level spectrum in the presence of a parabolic mass potential,
\begin{equation}
E_{m,n} = \tilde{v}_{m}^{+} \pm \sqrt{E_m^2+\left[\tilde{v}_m^-
+\mathrm{sgn}(\eta^-)\sqrt{\gamma^{-}}(2n+1)\right]^2}.
\label{energy2}
\end{equation}
The energy dependence with respect to the second discrete number $n$ is now quite
different from the previous Fock-Darwin-type spectrum in a scalar 2D-parabolic
potential, Eq.~(\ref{energy1}).
As a specific illustration for the case of a circular parabolic mass potential
$V_z({\bf r}) = (1/2) U_0 (x^2+y^2)$ together with a zero scalar term
$V_s=0$, the discrete energy levels are clearly anharmonic with respect to $n$,
\begin{equation}
E_{m,n} = -\frac{l_B^2}{2} U_0 \pm \sqrt{(\hbar \Omega_c \sqrt{m})^2
+ [l_B^2 U_0 \left(m+n+1/2\right)]^2},
\end{equation}
an expression which was already quoted in Eq. (\ref{spectrum2}).

\subsection{Discussion for arbitrary smooth potentials: A hierarchy of local energy scales}
\label{hier}

It is worth emphasizing that for arbitrary two-dimensional potentials $V_{s}({\bf R})$
and $V_{z}({\bf R})$ that are smooth at the scale of the magnetic length
$l_{B}$, the present vortex formalism turns out to be extremely useful
because it explicitly puts forward the existence of a hierarchy of local energy
scales. Such a hierarchy can then be exploited to devise successive
approximation schemes, leading to controlled expressions for all physical observables
at finite temperature.
This has already been proved with the concrete example of the temperature-broadened
STS local density of states for the 2DEG (see Sec. IV of Ref. \onlinecite{Champel2009}),
and the same mechanism holds also in the case of graphene studied here.

To understand qualitatively the origin of this hierarchy of local
energy scales, it is useful to rewrite the $\star$-bidifferential operator, Eq. 
(\ref{star}), under the equivalent form,
\begin{eqnarray}
\star &=& \sum_{p=0}^{+ \infty} \frac{1}{p!}\left(i\frac{l_{B}^{2}}{2} \right)^{p}
\hat{C}^{p} \label{op1}
\end{eqnarray}
with
\begin{eqnarray}
\hat{C}=\left(\overleftarrow{\partial}_{X}\overrightarrow{\partial}_{Y}
-\overleftarrow{\partial}_{Y}\overrightarrow{\partial}_{X} \right).
\end{eqnarray}
The arbitrary large number of derivatives in expression (\ref{op1}) is clearly an 
indication of the nonlocal nature of quantum mechanics. However, and remarkably here, we
realize that nonlocality manifests itself through {\em quasilocality} in
the vortex representation. This is due to the fact that the nonlocal electronic Green's 
function $\hat{G}({\bf r},{\bf r}')$ can entirely be determined from the
knowledge of the {\it local} vortex function $\tilde{g}_{m}({\bf R})$, see connection 
formula (\ref{passage2}). This quasilocality
property (which holds independently of the form of the potential landscape and thus
can be seen as resulting uniquely from the coherent character of the
vortex states), allows one to have a quasilocal quantization view. Clearly, the
local Green's function $\tilde{g}_{m}({\bf R})$ depends on the potential matrix
elements $\tilde{v}_{m}({\bf R})$ via the action of the
$\star$ product, see Eqs. (\ref{Dysondiag})-(\ref{Dyson0bis}). As obvious from
expression (\ref{op1}), each power of the bidifferential operator $\hat{C}$ acting on
the functions $\tilde{v}_{m}({\bf R})$ and $\tilde{g}_{m}({\bf R})$ generates
higher and higher derivatives $l_{B}^{p} \partial_{{\bf R}}^{p}
\tilde{g}_{m}({\bf R})$ of the local Green's function associated with hierarchy
of energy scales of the type $l_{B}^{p} \partial_{{\bf R}}^{p}
\tilde{v}_{m}({\bf R})$. These energy scales get smaller and smaller at
increasing $p$ in the case of a potential smooth at the magnetic length scale,
allowing one to control systematically the calculation.

For instance, leading order expressions (\ref{g1D0}) and (\ref{g1Dm}) for the
vortex Green's function were derived assuming that one can neglect potential
curvature terms (associated to the geometric invariants involving second-order
spatial derivatives of the potential).
This type of approximation is in fact {\it controlled} as long as temperature
exceeds the local energy scales appearing at next to leading order, respectively
$[\gamma({\bf R})]^{1/2}$ of Eq.~(\ref{gamma}) and $[\eta({\bf R})]^{1/3}$
of Eq.~(\ref{eta}).
In that case, quantum effects such as quantization and tunneling are certainly
missing, yet this basic approximation already encodes the structure of the delocalized
edge states far from the regions where the potential is strongly curved.

We have seen in Sec.~\ref{curved} that it is possible to go one step further
by including the curvature contributions [term $p=2$ in Eq. (\ref{op1})], and this
reintroduces quantization and tunneling in case of confined or open potentials,
respectively. Again, one expects that the refined expressions obtained for the vortex
Green's function [Eqs.~(\ref{g0}), (\ref{solution}), and (\ref{gVzquadra})
depending on the dominant type of scatterers] encode correctly the quantum
dynamics down to further and even smaller energy scales associated to geometrical
invariants involving third order spatial derivatives of the potential.

These considerations show the existence of a hierarchy of local energy scales
formed by the successive spatial derivatives of the potential and hint that the
passage from purely local physics (which is the hallmark of classical mechanics)
to highly nonlocal quantum-mechanical physics (which is the apanage of highly
unstable quantum states) is worked out {\em gradually} when the temperature is
progressively decreased. Therefore, at least in the large magnetic field regime,
it is not needed to diagonalize numerically the random Schr\"{o}dinger
or Dirac equation in order to calculate precisely physical quantities, since
temperature down to the Kelvin range in real experiments is not likely to be very
small compared to the tiny energy scales at order $l_B^2$ (for smooth
potentials). What is neglected in our approximation scheme are contributions
of some highly non-local quantum states superpositions, which are irrelevant in
realistic experiments at {\it finite} temperature.

\section{Connection with the deformation quantization theory}
\label{sec:deformation}

\subsection{Deformation quantization theory in classical phase space}

Before exploiting the expressions for the Green's functions derived in Sec.~\ref{high},
we would like to make important comments on the structure of the dynamical
equations obeyed by the Green's functions $\tilde{g}_{m}({\bf R})$ [general Dyson
Eq. (\ref{Dysonfinal}) at any magnetic field or Eq. (\ref{Dysondiag})
in the absence of Landau-level mixing at high magnetic field].
After completion of Ref. \onlinecite{Champel2009}, we have indeed realized that
the $\star$-bidifferential operator, Eq. (\ref{star}), involved in these latter
equations has a form analogous to the so-called star product, which has been
the subject of intense research in mathematical and in high-energy physics
because of its fundamental role in the principles themselves of quantum
mechanics. \cite{Bayen1978,Zachos2000,Zachos2002} More precisely, there have
been many attempts to formulate quantum mechanics from a classical point of
view, i.e., as a theory of functions on phase space, and one suggestion
\cite{Bayen1978} was to understand quantization as a deformation of the
structure of (Poisson-Lie) algebra of classical observables. The
$\hbar$-deformation theory of the classical mechanics relies on the
introduction of a star product,
\begin{eqnarray}
\star_{\hbar} =\exp\left[i \frac{\hbar}{2} \left( \overleftarrow{\partial}_{x}
\overrightarrow{ \partial}_{p_{x}}
- \overleftarrow{\partial}_{p_{x}} \overrightarrow{ \partial}_{x} \right) \right]
\label{star2}
\end{eqnarray}
in place of the usual product between phase-space functions. Here, $x$ and
$p_{x}$ are, respectively, the position and momentum which are canonically
conjugate variables. We discuss first here the quantization for a particle in
one dimension in the absence of a magnetic field (in two dimensions, classical
phase space is four dimensional, see discussion in Sec.~\ref{mixed}).
As a key principle, the entire quantum dynamics is encapsulated in
the noncommutative operator, Eq. (\ref{star2}), which turns out to be the unique
associative pseudodifferential deformation of the ordinary product.
Within the deformation quantization theory, the Poisson brackets of classical
mechanics between two phase-space functions $f(x,p_{x})$ and $g(x,p_{x})$ are
replaced by the Moyal brackets \cite{Moyal1949} defined as commutators (in
the star-product sense)
$[f,g]_{M}=(f \star_{\hbar} g-g \star_{\hbar} f)/i \hbar $. Obviously, Moyal
brackets are $\hbar$-dependent brackets which reduce smoothly to the Poisson
brackets in the limit $\hbar \to 0$ (hence the origin of the ``deformation''
picture).

The deformation quantization approach appears as a generalization of original
ideas put forward by Weyl, Wigner and Moyal \cite{Moyal1949} (for a short
historical account, see paper\cite{Zachos2002} and references therein),
which were aimed at getting a sound insight into the correspondence
principle between classical and quantum mechanics. The deformation quantization
formulation has acquired a clearer mathematical status 30 years ago with the
work of Bayen {\em et al.},  \cite{Bayen1978} where its autonomous and alternative
character with respect to other formulations of quantum mechanics, such as the
conventional Hilbert space and path integral formulations, has been proved (for
the recent status of the theory, see Refs. \onlinecite{Zachos2002} and \onlinecite{Hirshfeld2002}).
Because the basic {\em continuous} structure of the classical phase space is
conceptually kept in the deformation quantization theory, classical
mechanics is easily and transparently recovered via a smooth transformation, in
full contrast to the conventional operatorial approach of quantum mechanics
formulated in a Hilbert space (spanned by a {\em countable} basis of square
integrable states) where the emergence of a classical character from the quantum
substrate appears singular and rather challenging. For this reason, it has been
underlined \cite{Bayen1978} that the deformation view is presumably the right
way to look at quantization.

\subsection{Vortex Green's functions as a mixed phase-space formulation of
quantum mechanics}
\label{mixed}

Now considering explicitly two-dimensional electronic quantum dynamics in the ordinary 2DEG, the standard
deformation quantization theory introduces electronic coordinates $(x,y)$
and momenta $(p_x,p_y)$ as natural variables in a four-dimensional phase space.
In a large magnetic field however, the electronic classical dynamics consists
of a fast cyclotron motion, which is centered around a slowly moving guiding
center ${\bf R}=(X,Y)$.
In the popular operatorial language of quantum mechanics, these two relevant
degrees of freedom are introduced by decomposing the electronic coordinate
operator $\hat{{\bf r} }=\hat{{\bf R}} +\hat{{\bm \eta}}$ into a relative
position $\hat{{\bm \eta}}$ linked to cyclotron orbits and a guiding center
position $\hat{{\bf R}} =(\hat{X},\hat{Y})$. It is well known that the guiding
center coordinate operators obey the commutation rule
$[\hat{X},\hat{Y}]=il_{B}^{2}$, showing analogy with the canonical quantization
rule between the position $\hat{x}$ and the conjugate momentum $\hat{p}_{x}$.
Therefore, the square of the magnetic length, $l_{B}^{2}$, plays the role
of an effective magnetic field-dependent Planck's constant. Moreover, cyclotron
motion associated to the relative circular orbits $\hat{{\bm \eta}}$ leads to
quantized Landau levels and at very large magnetic fields completely decouples
from the guiding center dynamics.

This physical discussion shows that the canonical description of phase space in terms
of electronic coordinates $(x,y)$ and momenta $(p_x,p_y)$ becomes
awkward in a magnetic field. Quantum mechanically, this is reflected by the property
that states that are coherent with respect to both positions and momenta \cite{Feldman1970,Varro1984} cannot
be eigenstates of the kinetic part of the Hamiltonian associated to cyclotron motion,
contrary to the vortex states.
With the benefit of hindsight, the program that we have followed in the string
of recent papers\cite{Champel2007,Champel2008,Champel2009} is precisely
the formulation of deformation quantization in a {\it mixed} phase space associated
with the combination of discrete Landau levels $m$ and two-dimensional guiding center
coordinates $(X,Y)$, which correspond to physical space. For the 2DEG, this decomposition is naturally 
encoded within the vortex
states $\Psi_{m,{\bf R}}$ of Eq. (\ref{Sol2}), whose coherent character with respect
to the guiding center ${\bf R}$ brings a doubly continuous parametrization of
phase space, while the discrete quantum number is associated to a standard quantization
of cyclotron motion.

The general equation of motion at any magnetic field for graphene is then given by Eq. (\ref{Dysonfinal}),
and simplifies into a dynamics in two-dimensional phase space $(X,Y)$ given by Dyson
Eq. (\ref{Dysondiag}) in the large magnetic field regime, as cylotron motion giving rise to Landau levels exactly
decouples from the vortex (or guiding center) motion. In that case, Dyson equation has precisely the form
of a star product, see the obvious connection between the $\star$ operator, Eq.  (\ref{star}),
of the vortex formalism and the $\star_{\hbar}$ product, Eq. (\ref{star2}), of the deformation
quantization theory. High magnetic field dynamics is thus isomorphous to a one-dimensional
Schr\"{o}dinger (for the ordinary 2DEG) or Dirac equation (for graphene) with conjugate variables $X$ and $Y$. More
specifically, if we consider the lowest Landau level (allowing one to forget the
spinorial structure proper to graphene), Dyson Eq. (\ref{Dysondiag}) is
equivalent to the standard operatorial formulation with the Hamiltonian
$H = \tilde{v}_0(\hat{X},\hat{Y})$, where the effective potential $\tilde{v}_0({\bf R})$
is given by Eq.~(\ref{tildev0}). In that case, dynamics results from the commutation
rule $[\hat{X},\hat{Y}]=il_{B}^{2}$ so that kinetic-like energy terms emerge
from the identification of the conjugate momentum to $\hat{X}$ with
$\hat{P}_X= \hbar \hat{Y}/l_B^2$.
We emphasize that this derivation is free of the ambiguities found in the
path integral formulation  \cite{Jain} and reproduces the lowest Landau
projection method pioneered for the 2DEG by Girvin and Jach. \cite{Girvin,Jain} The
vortex formulation of phase space is however more general, because it
allows to consider not only the projection onto arbitrary Landau levels
at infinite magnetic field, but also the coupling between them for arbitrary
magnetic field.


Therefore, the semicoherent character of the vortex representation offers a
{\it local} quantization view in high magnetic field because phase space reduces
to the {\it physical} space of guiding center coordinates ${\bf R}$. When considering
the motion in complicated potential landscapes, this leads to the existence of a
hierarchy of local energy scales, allowing one to describe smoothly the crossover
from the semiclassical guiding center motion at high temperature to the fully quantum
dynamics at very low temperature, as discussed in Sec.~\ref{hier}.



\section{Local density of states}
\label{sec:local}

\subsection{Generalities}
We now use the formalism developed in the previous sections and the resulting
expressions for the graphene Green's function to investigate the characteristic
features of the local density of states (LDoS). The goal of Sec.
\ref{sec:local} is to show that a lot of information concerning the different
potentials at play in graphene can be extracted from the widths and shapes of
the LDoS peaks in a high magnetic field.

The LDoS is related to the electronic Green's function via
the formula
\begin{eqnarray}
\rho({\bf r},\omega) &=&- \frac{1}{\pi} \mathrm{Im}\, \mathrm{Tr} \, \hat{G}({\bf r},{\bf r},\omega) .
\label{defdos}
\end{eqnarray}
Note that with Eq. (\ref{passage2}), we can directly write the LDoS in terms of the modified local Green's function
$\tilde{g}_{m_{1},\lambda_{1};m_{2},\lambda_{2}}({\bf R})$. In the case where
the modified Green's function is diagonal with respect to the Landau-level
quantum number, i.e., $\tilde{g}_{m_{1},\lambda_{1};m_{2},\lambda_{2}}({\bf R})=
\delta_{m_{1},m_{2}} \, \tilde{g}_{m_{1};\lambda_{1};\lambda_{2}} ({\bf R})$, we
have to evaluate the action of the exponential differential operator onto the
product of two vortex functions with identical Landau level, as done in
Eq.~(\ref{Km}).
We therefore find that the LDoS [Eq. (\ref{defdos})] can 
quite generally be written in the absence of Landau level mixing as 
\begin{eqnarray}
\rho({\bf r},\omega) &=& 
- \frac{4}{ \pi} \mathrm{Im}
\int \!\!\!
\frac{d^{2}{\bf R}}{2 \pi l_{B}^{2}}
\Big[K_{0}({\bf R}-{\bf r}) \tilde{g}_{0}({\bf R})
+\frac{1}{2}
\sum_{m=1}^{+ \infty} \sum_{\lambda_{1},\lambda_{2}}
\nonumber \\
&& \hspace{-1cm} 
\times
\left\{
\lambda_{1} \lambda_{2} K_{m-1}({\bf R}-{\bf r}) + K_{m}({\bf R}-{\bf r})
\right\}
\tilde{g}_{m;\lambda_{1};\lambda_{2}}({\bf R}) \frac{}{}
\Big],
\nonumber
\\
\label{rhosimpler}
\end{eqnarray}
where the kernel $K_m({\bf R})$ has been previously obtained in Eq.~(\ref{Km}).
We have also taken into account here the spin and valley degeneracies, which provide
an overall prefactor of 4 when evaluating the trace in formula (\ref{defdos}).

In actual experimental conditions, one never has a direct access to the zero-temperature
LDoS, due to an extrinsic smearing occasioned by the finite
temperature $T$. The STS spectra at fixed energy $\varepsilon$ are proportional
to the temperature broadened LDoS
\begin{eqnarray}
\rho^{STS}({\bf r},\varepsilon,T)
= -\int d \omega \rho({\bf r},\omega) n_F'(\omega-\varepsilon),
\label{rhoSTS}
\end{eqnarray}
where $n_F'(\omega)=-1/[4 T \cosh^{2}(\omega/2T)]$ is the derivative
of the Fermi-Dirac function.

\subsection{LDoS for locally flat potentials}

\subsubsection{General expression}
The leading order result for the vortex Green's function, Eqs. (\ref{g1Dm}) and (\ref{g1D0}),
applies when the disorder potential is locally flat on the scale $l_B$. 
Mathematically, this approximation is controlled for temperatures larger than
the smaller energy scales associated to local Gaussian curvature, such as 
Eq.~(\ref{gamma}). In that case, using previous formulas (\ref{rhosimpler}) and 
(\ref{rhoSTS}), we get
\begin{eqnarray}
\nonumber
\rho^{STS}({\bf r},\varepsilon,T)
= -4
\int \!\!\! \frac{d^{2}{\bf R}}{2 \pi l_{B}^{2}}
\Big[n_F'(\varepsilon-\xi_{0}({\bf R}+{\bf r})) \,
K_{0}({\bf R})
\nonumber 
\\
 +\frac{1}{2} \sum_{m=1}^{+ \infty} \sum_{\epsilon=\pm} n_F'(\varepsilon-\xi_{m,\epsilon}({\bf R}+{\bf r}))
\left\{ \left(1+\epsilon \beta_{m}({\bf R}+{\bf r}) \right)
\right. \nonumber 
\\
\left.
\times
K_{m}({\bf R}) +\left(1-\epsilon \beta_{m}({\bf R}+{\bf r})\right)
 K_{m-1}({\bf R})
\right\} 
\frac{}{}\Big], \hspace*{0.5cm}
\label{rhoFlat}
\end{eqnarray}
where the effective energy $\xi_{m,\pm}({\bf R})$, the electron-hole asymmetry parameter
$\beta_m({\bf R})$ and the kernel $K_m({\bf R})$ are given respectively by 
Eqs.~(\ref{effective}), (\ref{beta}) and (\ref{Km}). The kernels $K_m({\bf R})$
are oscillating yet normalized functions that are localized around ${\bf R}={\bf 0}$ on a
characteristic length scale $L_{m} = l_B \sqrt{2m+1}$, which one associates with
the cyclotron radius. 
Only for the lowest Landau level $m=0$ does this length reduce to the magnetic length $l_{B}$. 

In principle, one cannot strictly set the temperature to zero in Eq. (\ref{rhoFlat}) unless the
effective potentials $\tilde{v}_{m}^{+}({\bf R})$ and $\tilde{v}_{m}^{-}({\bf
R})$ which compose the function $\xi_{m,\epsilon}({\bf R})$ are {\em globally}
flat. Indeed, for arbitrary potentials $\tilde{v}_{m}^{+}({\bf R})$ and
$\tilde{v}_{m}^{-}({\bf R})$, it is important to have in mind that expression
(\ref{rhoFlat}) overlooks the fine structure of the zero-temperature local density of
states, which requires to take into account all existing spatial derivatives of
these potentials $\tilde{v}_{m}^{\pm}({\bf R})$ [see Eq. (\ref{Dysondiag})].
Nevertheless, it captures accurately the shape of the LDoS when the temperature
exceeds the (smaller) energy scales involving second and higher derivatives (in
orthogonal directions) of the potentials $\tilde{v}_{m}^{\pm}({\bf R})$
associated to curvature. 
Basically, under the inequalities $L_{m}|{\bm \nabla} \xi_{m,\epsilon}({\bf r})|
> T \gg \sqrt{|\gamma_{m}^{\pm}({\bf R})|}$, one expects that the temperature
gives a small contribution to the smearing of the LDoS in comparison to the
intrinsic smearing generated by the spatial dispersion of the function
$\xi_{m,\epsilon}({\bf R}+{\bf r})$, (i.e., by the potential gradients) when
performing the integration over the vortex position ${\bf R}$ in Eq.
(\ref{rhoFlat}).  

\subsubsection{High-temperature regime}
At very high temperatures such that $T\gg L_{m}
|{\bm \nabla} \xi_{m,\epsilon}({\bf r})|$, the spatial dependence on the vortex position
${\bf R}$ inside the Fermi derivative function can be neglected [here, we also disregard the 
${\bf R}$ dependence of the smooth function $\beta_{m}({\bf R}+{\bf r})$], so that
expression (\ref{rhoFlat}) simplifies into:
\begin{eqnarray}
\label{rhoHT}
\rho^{STS}({\bf r},\varepsilon,T) & = & 
\frac{(-4)}{2\pi l_B^2}\\
\nonumber
&&\hspace{-1cm}\times \left[ n_F'(\varepsilon-\xi_{0}({\bf r})) +
\sum_{m=1}^{+ \infty} \sum_{\epsilon=\pm}
n_F'(\varepsilon-\xi_{m,\epsilon}({\bf r}))
\right].
\end{eqnarray}
This semiclassical expression provides LDoS peaks of width $2T$ that are
centered around the effective Landau-level energies $\xi_{m,\pm}({\bf r})$ given by Eq.~(\ref{effective}).
In this regime, the thermal broadening of the LDoS peaks is thus independent of the Landau-level
index, and the electron and hole peaks are characterized by the same heights.
At lower temperatures, we now show that different linewidths, line shapes and particle-hole
asymmetries are generated in the LDoS spectra, providing additional insight into
the underlying scalar and mass potentials.

\subsubsection{Low-temperature regime for potentials smooth on the cyclotron radius $L_{m}$} 

In case when $T\leq L_{m} |{\bm \nabla} \xi_{m,\epsilon}({\bf r})|$, the spatial
dependence of the Fermi function derivative must be kept. 
We first assume here that the potential is well approximated by its first-order
gradient on the {\it whole} cyclotron orbit of radius $L_{m}$,
i.e., $\xi_{m,\epsilon}({\bf R}+{\bf r})\simeq \xi_{m,\epsilon}({\bf r}) +{\bf R}.{\bm \nabla}
\xi_{m,\epsilon}({\bf r})$. We can then perform analytically the Gaussian integral over ${\bf R}$ 
in Eq.~(\ref{rhoFlat}) and obtain the intuitive result for the zero-temperature
LDoS (see Appendix~\ref{appD}),
\begin{widetext}
\begin{eqnarray}
\rho({\bf r},\omega) \simeq
\frac{1}{2 \pi l_{B}^{2}} \frac{4}{\sqrt{\pi}} 
\left[ 
\frac{1}{\Gamma_0^{loc}({\bf r})}
\exp\left\{-\left(\frac{\omega-\xi_{0}({\bf r})}{\Gamma_{0}^{loc}({\bf r})}
\right)^{2}\right\}
+ \frac{1}{2} \sum_{m=1}^{+ \infty} \sum_{\epsilon=\pm} 
\frac{1}{\Gamma_{m,\epsilon}^{loc}({\bf r})} 
\left\{ 
\frac{1+\epsilon \beta_{m}}{2^m m!}
H_m^2
\left[
\frac{\omega-\xi_{m,\epsilon}({\bf r}) }{\Gamma_{m,\epsilon}^{loc}({\bf r})} \right]
 \nonumber \right. \right. \\
\left. \left.
+ \frac{1-\epsilon \beta_{m}}{2^{m-1} (m-1)!}
H_{m-1}^2
\left[
\frac{\omega-\xi_{m,\epsilon}({\bf r}) }{\Gamma_{m,\epsilon}^{loc}({\bf r})} \right]
\right\}
\exp\left\{-\left(\frac{\omega-\xi_{m,\epsilon}({\bf r})}
{\Gamma_{m,\epsilon}^{loc}({\bf r})} \right)^{2}\right\}
\right] ,
\label{dosgradient}
\end{eqnarray}
\end{widetext}
with $\Gamma_{m,\epsilon}^{loc}({\bf r})=l_{B} |{\bm \nabla} \xi_{m,\epsilon}({\bf r})|$
the {\it local} energy scale associated to the drift motion and $H_m(x)$ the $m$th 
Hermite polynomial. In order to keep
the above expression compact, we have only written the zero-temperature
local density of states, but the STS local density of states is readily obtained from
Eq.~(\ref{rhoSTS}).
The above expression is quite reminiscent of the expression that can be obtained 
with the usual Landau states, of course
generalized to the two-component spinorial structure proper to graphene,
and taking into account that the potential landscape varies slowly in space
[obvious from the ${\bf r}$ dependence of the width 
$\Gamma_{m,\epsilon}^{loc}({\bf r})$]. In Sec.~\ref{experiment}, we will further 
analyze expression~(\ref{dosgradient}) when discussing recent STS experiments. 

We note yet that for a given disordered potential landscape 
Eq.~(\ref{dosgradient}) breaks down for sufficiently large quantum 
numbers $m$, because very wide cyclotron orbits of radius $L_m \approx \sqrt{2m}l_B \gg l_{B}$ may
explore random spatial variations in the potential. In that case, the more general
expression~(\ref{rhoFlat}) is still valid, provided that the potential is smooth on
the smaller scale $l_B$ (this is always the case at high enough magnetic field).
This regime is now investigated.

\subsubsection{Low-temperature regime for potentials with random spatial 
fluctuations on the cyclotron radius $L_{m}$}

In cases where the disorder potential fluctuates spatially on the scale of cyclotron radius
$L_{m}$, formula~(\ref{dosgradient}) is clearly invalid, as the linearization
of the effective vortex potential $\xi_{m,\epsilon}({\bf{R}+\bf{r}})$ cannot
be made anymore. When spatial variations along the trajectory remain however
smooth at the smaller scale $l_B$, general expression~(\ref{rhoFlat}) for locally flat
potentials is the one to consider. In order to get some analytical insight, we
compute here a {\it disorder averaging} of the LDoS. This procedure is clearly
valid in two cases: (i) for the LDoS at very large Landau index $m\gg1$, 
as very wide cyclotron radius $L_{m}$ can explore random configurations of the scalar 
disordered potential $V_{s}({\bf r})$. Because of the large quantum numbers
involved here, one should recover a semiclassical limit, as we will see;
(ii) for any $m$ and finite magnetic length (the fully quantum regime), if one rather 
considers the sample averaged density of states (DoS). We stress beforehand that
the LDoS at small $m$ does {\it not} show self-averaging.
In both situations, the computed averaged density of states is a spatial-independent
quantity. The calculation performed in Appendix~\ref{appD} provides the following result:
\begin{eqnarray}
\label{average}
\rho^{DoS}(\omega) &\equiv& \overline{\rho({\bf r},\omega)} \\
\nonumber
&=& \frac{1}{2\pi l_B^2}  \frac{4}{\sqrt{\pi}}
\left[ \frac{1}{\Gamma_{0}^{DoS}} 
\exp\left\{-\left(\frac{\omega}{\Gamma_0^{DoS}}\right)^2\right\} \right.\\
\nonumber
&&\left.\hspace{-0.4cm}+\sum_{m=1}^{+\infty}
\frac{1}{2} \sum_{\epsilon=\pm}\frac{1}{\Gamma_{m}^{DoS}}
\exp\left\{-\left(\frac{\omega-\epsilon E_m}
{\Gamma_m^{DoS}}\right)^2\right\} \right], \hspace*{0.5cm} 
\end{eqnarray}
with the characteristic energy width $\Gamma_{m}^{DoS}$ given by
\begin{eqnarray}
\left[\Gamma_m^{DoS}\right]^2 = \int \!\!\! \frac{d^2 {\bf q}}{(2\pi)^2} 
2 S(q) \left| \int d^2{\bf r} \, e^{i{\bf q}.{\bf r}} 
\frac{K_m({\bf r})+ K_{m-1}({\bf r})}{2}
\right|^2 ,
\nonumber \\
\label{width}
\end{eqnarray}
where $S(q)$ is the Fourier transform of the potential correlation function (see
Appendix~\ref{appD} for details) and $K_m({\bf R})$ was defined in Eq.~(\ref{Km})
(we write here $K_{-1}\equiv K_0$ in order for the above formula to apply at 
$m=0$ as well).
In order to simplify the derivation, we have assumed that the antisymmetric part 
$V_z$ of the total potential $V$ can be neglected compared to the diagonal scalar component $V_s$.

Equation~(\ref{width}) can be first analyzed in the following semi-classical limit, 
$l_B\to0$ and $m\gg1$, while keeping the cyclotron radius $L_m=\sqrt{2m+1}l_B$ fixed.
In that case, the function $K_m({\bf r})$, which is peaked at 
the distance $|{\bf r}|=\sqrt{2m} l_B\simeq L_m$ with a width $l_B$, becomes a 
delta function along the cyclotron radius, $K_m({\bf r})\simeq \frac{1}{2\pi L_m} 
\delta(|{\bf r}|-L_m)$.
In this semi-classical regime, we recover results derived by other means~\cite{Raikh1993} 
for the 2DEG, namely,
\begin{equation}
\label{newwidth}
[\Gamma_m^{DoS}]^2 = \int \frac{q dq}{2\pi} 
2 S(q) |J_0(L_m q)|^2 \simeq \frac{\int dq\, 2 S(q)}{\pi^2 L_m}
\propto \frac{1}{\sqrt{m}},
\end{equation}
where the asymptotic limit of the zeroth order Bessel function $J_0$ was used,
assuming the disorder to be random on the scale $L_m$, so that
the integral in Eq.~(\ref{newwidth}) is dominated by its tail. We note
that our expression~(\ref{width}) is more general than the above result (\ref{newwidth}), because
it also describes the averaged density of states for any $m$ (including the strong quantum
regime at finite $l_B$). Clearly, our calculation incorporates wave function
spreads on the scale $l_B$, a purely quantum length scale which has completely disappeared 
from the semiclassical result [Eq.~(\ref{newwidth})]. In all cases (semiclassical
or quantum dynamics), the general trend is that the cyclotron motion averages
out the local potential at increasing radius $L_m$, so that the width of the DoS
decreases with $m$. This effect is discussed now in more detail at the light of recent 
LDoS measurements.

\subsection{Interpretation of the STS experiments}
\label{experiment}

Recent experimental works by Li {\em et al.} \cite{Li2009} and Miller {\em
et al.} \cite{Miller2009} have investigated by STS the LDoS in
graphene at high magnetic field and have revealed the relativistic nature of
the Landau levels in the measured energy spectrum.
Besides this precise verification of the sequence of graphene
Landau levels at the energies $\pm \hbar \Omega_c \sqrt{m}$, one can note several other striking 
aspects of the data. At a given large magnetic field and for a fixed tip position, 
the width of the $m$th Landau-level peak in the LDoS $\rho^{STS}(\varepsilon,{\bf r})$
is seen to {\it grow} as $\sqrt{m}$ with increasing Landau-level index $m$, as demonstrated in Ref.~\onlinecite{Li2009}
and also observed in Ref.~\onlinecite{Miller2009}. At the same time, 
the LDoS peaks display an energy dispersion as a function of tip position, reflecting
the underlying effective potential, see the discussion in Ref.~\onlinecite{Miller2009}. 
Quite contrary to the fixed tip LDoS peaks, the energy spread of the spatially averaged
$m$th Landau level {\it decreases} with $m$. This effect is easily
understood on general grounds by the smearing of the local potential by larger
and larger cyclotron orbits, as discussed above and embodied in the DoS width
$\Gamma_m^{DoS}$ of Eq.~(\ref{width}). In particular, the semiclassical limit ($m\gg1$, $l_B\to0$),
which does not completely apply to the experiment for which Landau levels are
only observed up to $m=7$, gives the result $\Gamma_m^{DoS}\propto m^{-1/4}$, as first 
derived by Raikh and Shahbazyan\cite{Raikh1993} for the non-relativistic 2DEG,
showing a clear decrease in the width with $m$.

\begin{figure}
\includegraphics[scale=1.19]{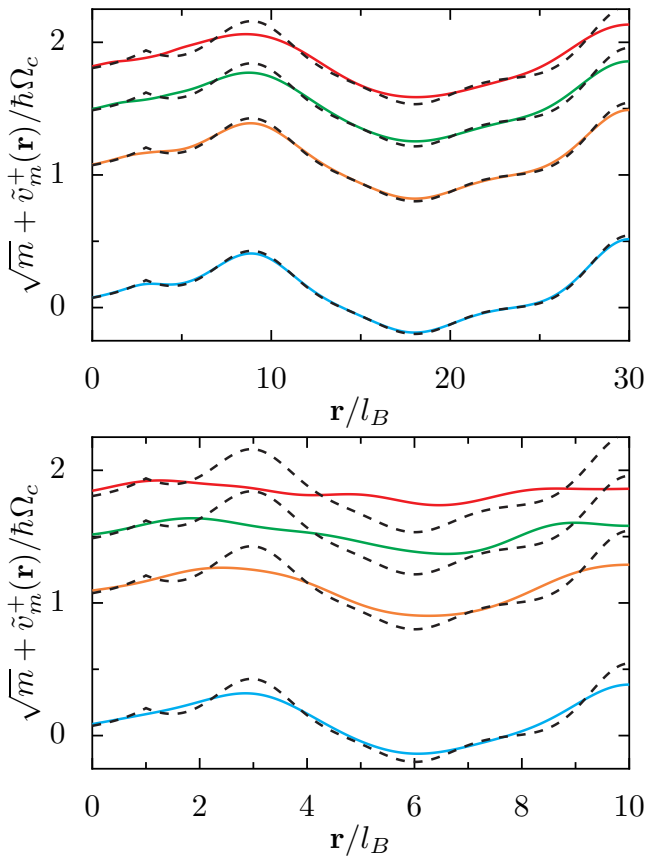}
\caption{(Color online) Dimensionless effective potential 
$\xi_{m,+}({\bf r})/\hbar\Omega_c=\sqrt{m}+\tilde{v}_m^+({\bf r})/\hbar\Omega_c$
from Eq.~(\ref{xieff})
as a function of linear tip position ${\bf r}/l_B$ for the first Landau levels $m=0,1,2,3$ (bottom 
to top in full lines), and compared to the bare potential energy 
$\sqrt{m}+V_s({\bf r})/\hbar\Omega_c$ given by the dashed lines.
The top panel corresponds to smooth disorder while the bottom one has stronger variations in
the potential on the scale $l_B$ (see the relative axes).}
\label{figVeff}
\end{figure}

As an illustration of truly {\it quantum} smearing of the cyclotron motion at finite
$l_B$ for the first few Landau levels, which corresponds more to the actual experimental
situation at high magnetic fields, we have plotted in Fig.~\ref{figVeff} for $m<4$ the 
effective potential in graphene obtained from Eqs.~(\ref{tildev}) and~(\ref{effective}) in the case of negligible band mixing [i.e.,
$|\tilde{v}^-_m({\bf r})|\ll\hbar\Omega_c$],
\begin{equation}
\xi_{m,+}({\bf r}) 
= E_m + \frac{1}{2} \int d^2 {\bm \eta} V_s({\bm \eta}) 
[ K_m({\bf r}-{\bm \eta}) + K_{m-1}({{\bf r}-\bm \eta})]
\label{xieff}
\end{equation}
as a function of tip position ${\bf r}$ and for a given (scalar) 
disorder realization, obtained as a superposition of localized long-range
potentials.
The upper panel of Fig.~\ref{figVeff}, which corresponds to a (uni-dimensional) disordered scalar potential landscape $V_{s}({\bf r})$ smooth on the scale $l_B$,
shows that the effective potential $\xi_{m,+}({\bf r})$ follows precisely the bare disorder
potential for the lowest Landau level $m=0$, yet presents some moderate
deviations for the following levels, illustrating the small averaging
present on the larger scale of the quantum cyclotron radius $L_m=\sqrt{2m+1}l_B$. 
In contrast, the lower panel of Fig.~\ref{figVeff} presents the situation of a disordered scalar potential landscape $V_{s}({\bf r})$ which has
spatial variations comparable to $l_B$ [we stress again that the effective potential
given by Eq.~(\ref{effective}) and thus also by Eq. (\ref{xieff}) has a truly non-perturbative character in $l_B$].
In that case, we can notice two effects:
(i) the effective potential $\xi_{m,+}({\bf r})$ shows important quantitative deviations from the bare one
already in the lowest Landau level $m=0$; (ii) at increasing $m>0$, stronger
and stronger averaging effects take place, so that the effective potential $\xi_{m,+}({\bf r})$
rapidly flattens out. As a consequence, the typical energy width of the
effective potentials $\xi_{m,\pm}({\bf r})$ as a function of position ${\bf r}$
clearly {\it decreases} with growing $m$. This effect is clearly seen in
the STS data of Ref.~\onlinecite{Miller2009} for graphene and can be also
recognized in recent measurements on standard 2DEGs by Hashimoto {\it et 
al}.\cite{Hash2008}

We now discuss in more detail the STS spectra taken at {\it fixed} tip
position, presented in the experimental papers~\cite{Li2009,Miller2009}
that showed a {\it broadening} of the Landau levels with a $\sqrt{m}$ scaling at increasing $m$.
At high temperatures, such that $T\gg L_{m} \left|{\bm \nabla} \xi_{m,\pm}({\bf r})\right|$, the
broadening has a purely thermal origin, with a fixed width set by $T$ and an
exponential line shape (given by the Fermi function derivative). It is worth noting that the apparent
increase with $m$ of the heights of the LDoS peaks in
graphene~\cite{Li2009,Miller2009} is solely due to the collapse
of Landau levels, $E_{m+1}-E_m \propto 1/\sqrt{m}$ at large $m$, yet the
underlying Landau peaks show a width insensitive to $m$.

\begin{figure}
\includegraphics[scale=1.19]{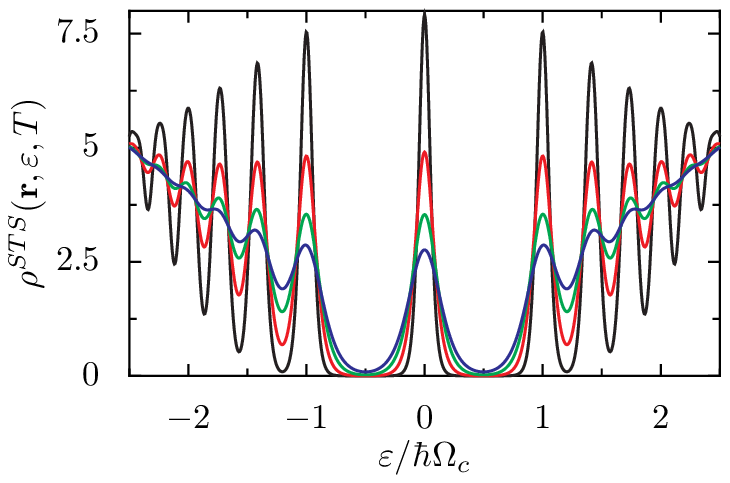}
\caption{(Color online) Energy-dependent STS spectra for the local density of states
$\rho^{STS}({\bf r},\varepsilon,T)$ at {\it fixed} tip position ${\bf r}$ 
from Eqs.~(\ref{rhoSTS}) and~(\ref{dosgradient}) in units of $4/(2\pi l_{B}^{2})$ and as a function of
energy $\varepsilon$ for several temperatures
$T/\hbar\Omega_c=0.03,0.05,0.07,0.09$ (top to bottom). Here the local energy
scale associated to the drift motion  in the lowest Landau level is $\Gamma_0^{loc}({\bf r}) = l_B |{\bm \nabla} V_s| 
= 0.02 \hbar \Omega_c$. At the lowest given temperature, the smearing with increasing $m$
of the Landau-level peaks is characterized by a local energy width which roughly grows as $\sqrt{m}$
(thermal smearing provides still some dominant contribution).}
\label{figDOS}
\end{figure}

Contrary to the discussion given in Ref. \onlinecite{Li2009}, we emphasize
that results of disorder averaged density of states, such as our Eq.~(\ref{newwidth}) or 
the formula obtained, e.g., in Ref.~\onlinecite{Peres2006}, do {\it not} apply to account for the 
width of the STS peaks at fixed position, for which an expression for the {\it local} density
of states, such as Eq.~(\ref{rhoFlat}) or ~(\ref{dosgradient}), should instead be 
considered. In fact, the energy spread of the Landau-level peaks observed experimentally at low
temperature in the LDoS can be easily understood to originate from 
wave-function broadening.
Indeed, in formula (\ref{dosgradient}) for instance, the polynomial $|H_m(x)|^2$ being of order $2m$, the squared wave function
$f(x)=|H_m(x)|^2 e^{-x^2}$ turns out to be spread on a characteristic scale
$x_m=\sqrt{2m+1}$. We note that in Fig. 2 of Ref.~\onlinecite{Kramer2010}, a square-root
dependence of the Landau-level widths with the Landau-level index can also be
observed at zero-temperature (the oscillatory substructure of each Landau level
peak disappears when including a small thermal smearing, as performed here).
Turning to the LDoS expression~(\ref{dosgradient}), one sees that 
the effective energy width of the $m$th Landau-level peak is roughly given by the
{\it local} energy scale $\sqrt{2m+1} l_B 
|{\bm \nabla} V_{s}({\bf r})| =  L_m |{\bm \nabla }V_{s}({\bf r})|$ [here we have used the fact 
that the effective potential $\xi_{m,\pm}({\bf r})$ roughly follows the bare potential $V_{s}({\bf r})$], 
which scales as $\sqrt{m}$ as observed in the experiment. \cite{Li2009} 
This effect can be checked by a simple numerical evaluation of Eq.~(\ref{dosgradient}), taking 
into account the convolution with a thermal smearing as resulting from Eq.~(\ref{rhoSTS}) for the STS local density of states. The obtained result for  the sequence of LDoS peaks  is shown in Fig.~\ref{figDOS}  for different temperatures.
At temperatures comparable to $\Gamma^\mathrm{loc}_0({\bf r}) \sim l_{B}|{\bm \nabla} V_s|  $, quantum 
smearing due to the drift motion, which is encoded by the spatial dependence
of the kernel $K_m({\bf R})$ in the general expression~(\ref{rhoFlat}),
or by the Hermite wave functions in the special case of a globally flat
potential [see Eq.~(\ref{dosgradient})], starts to appear. The growth of the energy width of the 
LDoS peaks at increasing Landau-level index $m$ is visible for the lowest chosen temperature in Fig. \ref{figDOS}.
In that case, one also sees a clear {\it decrease} 
in the heights of the LDoS peak with $m$, as observed
experimentally.~\cite{Li2009,Miller2009,Hash2008} Because the total smearing of
the Landau levels depends both on thermal and intrinsic wave function broadening, the 
linewidth is only roughly behaving as $\sqrt{m}$.

Finally, we address the question of the Landau-levels line shape in the LDoS.
In experiment of Ref.~\onlinecite{Li2009}, it has been pointed out that Lorentzian fits are
significantly better than Gaussian ones to account quantitatively for the
broadening of the LDoS peaks. On the other hand, in experiment by Miller {\it et
al.} \cite{Miller2009} the line shape has been modeled by a convolution of Lorentzians and Gaussians to
include extrinsic origins of broadening induced by temperature and instrumental
resolution. On theoretical grounds, thermal broadening implies exponential
line shapes (in between Lorentzians and Gaussians), while intrinsic wave function broadening
of drift states (for non-vanishing local potential gradients) leads to Gaussian-type
energy decay. We also note that spectra taken in regions of small potential 
gradients involve intrinsic exponential linewidth due to curvature effects, see
Ref.~\onlinecite{Champel2009bis} for a discussion of the lowest Landau-level LDoS peak in 
the 2DEG. Therefore, it is difficult in general to disentangle the different contributions
from the experiment, and systematic studies in temperature and as function of
tip position, would be required to settle precisely this issue.

\section{Conclusion}

In this paper, we have extended to the graphene case a Green's-function
formalism well suited to study the mechanism of lifting of the Landau-level
degeneracy by a smooth potential landscape at high magnetic fields, which was
originally developed for the two-dimensional electron gases. The whole formalism
relies on the use of a particular representation of semicoherent states, which
are eigenstates of the kinetic part of the Hamiltonian. These so-called vortex
states in the 2DEG case, or graphene vortex states in the graphene case, are
both characterized by an integer topological quantum number $m$, related to the
vortex circulation and giving rise to the Landau quantization of the orbital
motion, and by a doubly continuous quantum number ${\bf R}$, corresponding to the
location of the vortexlike phase singularities of the electronic wave function
and characterizing the huge degeneracy of the Landau levels in the absence of
disorder. The coherent states
character with respect to the degeneracy quantum number ${\bf R}$ allows one to
project the electron dynamics onto this overcomplete representation of states,
which rigorously extends to quantum mechanics the classical guiding-center
picture.

In a first stage, we have derived the exact matrix elements for smooth arbitrary
scalar and mass potentials, as well as for off-diagonal smooth potentials
related to ripples in graphene. The particular form of these matrix elements has
revealed the different processes leading to Landau-level mixing and coupling
between electron and hole bands. We have shown that at high magnetic field, when
the Landau-level mixing can be safely neglected, a mixing between the hole and
electron energy bands is unavoidably induced by second-order derivatives of the
scalar potential, independently of the presence or not of a mass potential. We
have been able to derive in this high magnetic field regime exact expressions
for the electronic Green's function in the presence either of an arbitrary
quadratic scalar potential or an arbitrary quadratic mass potential.

Besides affording the derivation of unique Green's function solutions valid for
closed and open quadratic potentials which underline the dual correspondence
between quantization effects and tunneling effects, we have emphasized that the
semicoherent vortex representation offers a quasilocal perspective of the
quantization process closely related to the deformation view of the classical
phase-space mechanics, a property which turns out to be essential to capture the
transition from the nonlocal quantum world to the local classical world.
Furthermore, the vortex representation has revealed a hierarchy of local energy
scales formed by the successive derivatives of the potential and thus ordered by
their degree of nonlocality. As a result, quantum features associated with the
lowest derivatives of the potential appear to be the most robust against the
inelastic effects. We have emphasized that the consideration of a finite
temperature allows one to disregard the smallest inaccessible energy scales and
thus to devise successive approximation schemes for an arbitrary smooth 
potential.

Within this spirit, we have derived controlled analytical expressions for the
local density of states in graphene valid at high magnetic field in the presence
of smooth arbitrary scalar and mass potentials within different temperature
regimes. We have identified the most relevant mechanism of intrinsic broadening
of the LDoS peaks and have shown that a lot of information on the different
potentials at play in graphene can be extracted from the experimental LDoS
spectra performed at high magnetic field. Finally, we have been able to explain
a few of the experimental findings, e.g., concerning the scaling of the LDoS
peaks with the Landau-level index, recently observed \cite{Li2009,Miller2009} in 
scanning tunneling spectroscopy of graphene.

\section*{Acknowledgments}

We acknowledge interesting discussions with D.M. Basko, M.O. Goerbig, L. Magaud,
P. Mallet, and J.Y. Veuillen. We thank L. Canet for taking part in the early
stages of this work.

\appendix

\section{Energy spectrum for closed quadratic potentials \label{appC}}

In this appendix, we show how the energy spectrum for a confining quadratic
potential (with a positive Gaussian curvature) can be determined from a retarded
Green's function expression such as given, e.g., by Eq. (\ref{g0}). For
$\gamma_{0}({\bf R})>0$, the function
\begin{eqnarray}
W({\bf R},t)=\frac{e^{-i\left[\eta_{0}({\bf R})/\gamma_{0}({\bf R})\right]
\tau_{0}(t)}}{\cos\left(\sqrt{\gamma_{0}({\bf R})}t \right) }
\end{eqnarray}
is periodic in time with the period $T=2 \pi/\sqrt{\gamma_{0}({\bf R})}$ at fixed ${\bf R}$. 
We thus expand it in a Fourier series
\begin{eqnarray}
W({\bf R},t)=\sum_{p=-\infty}^{+ \infty} a_{p}({\bf R})
e^{-i p \sqrt{
\gamma_{0}({\bf R})}
t},
\label{Fourier1}
\end{eqnarray}
and insert expression (\ref{Fourier1}) into Eq. (\ref{g0}) to straightforwardly get after integration over time
\begin{eqnarray}
\tilde{g}_{0}({\bf R})=\sum_{p=-\infty}^{+\infty}
\frac{a_{p}({\bf R})}
{
\omega-w_{0}({\bf R})-p\sqrt{
\gamma_{0}({\bf R})}+i0^{+}
}
\label{g0trans}
\end{eqnarray}
with $w_{0}({\bf R})=\tilde{v}_{0}({\bf R})-\eta_{0}({\bf R})/\gamma_{0}({\bf R})$.

The Fourier coefficients $a_{p}({\bf R})$ are given by
\begin{eqnarray}
a_{p}({\bf R})& =& \frac{\sqrt{\gamma_{0}({\bf R})}}{2 \pi}
\int_{0}^{2 \pi/\sqrt{\gamma_{0}({\bf R})}}
 \!\!\!\!
dt \,
W({\bf R},t) \,
e^{i p \sqrt{\gamma_{0}({\bf R})}t} \hspace*{0.7cm}
\\
&=& \frac{\left[ 1-
(-1)^{p}
\right]}{2 \pi} \int_{-\pi/2}^{\pi/2} \!\!\!\! d\theta \, \frac{e^{-i\rho({\bf R}) \tan \theta }}{\cos \theta} \,
e^{ip \theta}
\label{theta}
\end{eqnarray}
with $\rho({\bf R})=\eta_{0}({\bf R})/\left[\gamma_{0}({\bf R}) \right]^{3/2}$. We rewrite the following function appearing in the integrand of integral (\ref{theta}) as
\begin{eqnarray}
 \frac{
e^{-i\rho({\bf R}) \tan \theta }}{\cos \theta} &= & 2
e^{\rho({\bf R})} \frac{e^{i\theta}} {1+e^{2i \theta}}
\exp \left[-2 \rho({\bf R}) \frac{e^{2 i \theta}}
{1+e^{2 i \theta}} \right]
\label{exp1}
 \\
&=& 2 e^{-\rho({\bf R})} \frac{e^{-i\theta}} {1+e^{-2i \theta}}
\exp \left[2 \rho({\bf R}) \frac{e^{-2 i \theta}}
{1+e^{-2 i \theta}} \right]
\nonumber
.
\\
\label{exp2}
\end{eqnarray}
It is then convenient to introduce the identity (see, e.g., Ref. \onlinecite{Gradstein})
\begin{eqnarray}
\frac{1}{z-1} \exp\left(\frac{ xz}{z-1} \right)
=\sum_{n=0}^{+ \infty} L_{n}(x) z^{n}
,
\label{identity}
\end{eqnarray}
where $L_{n}(x)$ is the Laguerre polynomial of degree $n$.
Formula (\ref{identity}) is usually defined for $|z|<1$, but it can be checked that it still holds for $z=e^{i \varphi} $ with $\varphi \neq 2 \pi j$ ($j$ a positive or negative integer) at $x>0$. Indeed, using the asymptotic behavior of the Laguerre polynomials at large $n$ and $x>0$,
\begin{eqnarray}
L_{n}(x) \approx \frac{e^{x/2}}{\sqrt{\pi} (nx)^{1/4}} \cos\left(2 \sqrt{x\left(n+\frac{1}{2}\right)} -\frac{\pi}{4} \right)
,
\end{eqnarray}
we note that the series on the right-hand side of Eq. (\ref{identity}) is semi-convergent (this can be established using Abel's test). On the other hand, for $x <0$, we have
\begin{eqnarray}
L_{n}(x) \approx \frac{e^{-x/2}}{2\sqrt{\pi} (n |x|)^{1/4}} \exp\left(2 \sqrt{ |x|\left(n+\frac{1}{2}\right)}\right)
,
\end{eqnarray}
meaning that the series on the right-hand side of Eq. (\ref{identity}) is divergent for $x<0$ and $z=e^{i \varphi} $.

Using Eqs. (\ref{exp1}) or (\ref{exp2}), and Eq. (\ref{identity}) by
writing  $x=2 |\rho({\bf R})|$ and $z=-e^{ -2 i \chi \theta }$ depending on the
sign of the quantity $\rho({\bf R})$ [we introduce the short-hand notation $\chi=\mathrm{sgn} \, \rho({\bf R})$], we can easily perform the integration over the angle $\theta$ in Eq. (\ref{theta}) and find
\begin{eqnarray}
a_{p}({\bf R})=2 (-1)^{n} e^{-\left|\rho({\bf R})\right|} L_{n}\left(2 \left|\rho({\bf R})\right| \right)
\end{eqnarray}
for $p=\chi(2n+1)$
and $a_{p}({\bf R})=0$ for any values of $p \neq \chi(2 n+1)$. Therefore, only
the terms with $p=\chi(2n+1)$ remain in expression (\ref{g0trans}), where $n$ is
a positive integer and $\chi=\pm1$ is an index determining if the region is locally convex or concave.

Now, for the particular case of purely quadratic scalar and mass potentials, the
poles of the Green's function (\ref{g0trans}) are ${\bf R}$ independent, and
thus directly yield the energy spectrum, Eq. (\ref{spectrumg0}), with the set of
quantum numbers $(m,n)$ if the quadratic potential $V_{s}-V_{z}$ is convex,
i.e., confining ($\chi=+1$ in this case).

\section{Solution for a locally quadratic scalar potential $V_{s}$ \label{A}}
\label{appA}

In this appendix, we solve the equations of motion, Eq. (\ref{Dysondiag}), in the regime where we can consider that the effective potential $\tilde{v}_{m}^{-}({\bf R})$ has a negligible spatial dispersion and that the effective
potential $\tilde{v}_{m}^{+}({\bf R})$ can be locally described up to its second-order derivatives (i.e., it is locally written as a two-dimensional quadratic potential). These assumptions turn out to be exactly fulfilled in the particular case of a globally quadratic scalar potential $V_{s}({\bf R})$ and a constant mass term $V_{z}$, for which 
the matrix elements at high magnetic field read for $m \geq 1$,
\begin{eqnarray}
\tilde{v}_{m}^{+}({\bf R})
&=&
V_{s}({\bf R})+m \frac{l_{B}^{2}}{2} \Delta_{{\bf R}} V_{s}({\bf R}), \\
\tilde{v}_{m}^{-}({\bf R})
&=&
-V_{z}+\frac{l_{B}^{2}}{4} \Delta_{{\bf R}} V_{s}({\bf R})= cst
.
\end{eqnarray}
Using the explicit form, Eq. (\ref{star}), of the $\star$ operator and the fact that
$\tilde{v}^{+}_{m}({\bf R})$ is a quadratic function (so that all its
derivatives higher than 3 vanish) and that $\tilde{v}^{-}_{m}({\bf R})$ is
quasi-independent of ${\bf R}$, Eq. (\ref{Dysondiag}) becomes
\begin{widetext}
\begin{eqnarray}
\left( \omega-\tilde{v}_{m}^{+}({\bf R})-E_{m,\lambda_{1}}+i 0^{+} \right) \,
\tilde{g}_{m;\lambda_{1};\lambda_{2}}({\bf R})
=
 \delta_{\lambda_{1},\lambda_{2}}
+ \tilde{v}_{m}^{-} \tilde{g}_{m;-\lambda_{1};\lambda_{2}}
({\bf R})+i\frac{l_{B}^{2}}{2} \left[ \partial_{X} \tilde{v}_{m}^{+} \partial_{Y}-
\partial_{Y} \tilde{v}_{m}^{\mathrm{+}} \partial_{X}
\right] \tilde{g}_{m;\lambda_{1};\lambda_{2}}({\bf R})
\nonumber
\\
-\frac{l_{B}^{4}}{8}
\left[ \left(
\partial_{Y}^{2}\tilde{v}_{m}^{+} \right) \partial_{X}^{2}
+
\left(
\partial_{X}^{2}\tilde{v}_{m}^{+}\right) \partial_{Y}^{2}
-2 \left(
\partial_{X} \partial_{Y}
\tilde{v}_{m}^{+}
 \right) \partial_{X}\partial_{Y}
\right]\tilde{g}_{m;\lambda_{1};\lambda_{2}} ({\bf R})
\label{DV0quadra1}
.
\end{eqnarray}
To solve Eq. (\ref{DV0quadra1}), we introduce an arbitrary reference point ${\bf R}_{0}$
and write $\tilde{g}_{m;\lambda_{1};\lambda_{2}}({\bf R})=f_{m;\lambda_{1};
\lambda_{2}}\left[E({\bf R}) \right]$ with $E({\bf R})=\tilde{v}_{m}^{+}({\bf R})-\tilde{v}_{m}^{+}({\bf R}_{0})$.
Substituting this form into Eq. (\ref{DV0quadra1}), we get the system of differential
equations,
\begin{eqnarray}
\left[ (\gamma_{m}^{+} E +\eta_{m}^{+}) \frac{d^{2}}{d E^{2}} 
+\gamma_{m}^{+} \frac{d}{dE}-E+\omega-\tilde{v}_{m}^{+}({\bf R}_{0})-E_{m,\lambda_{1}}+i0^{+}
\right]f_{m;\lambda_{1};\lambda_{2}} (E)
-
\tilde{v}_{m}^{-}
f_{m;-\lambda_{1};\lambda_{2}} (E)
=
 \delta_{\lambda_{1},\lambda_{2}}
, \nonumber \\
\label{DquadraV0}
\end{eqnarray}
where the geometric coefficients $\gamma_{m}^{+}$ and $\eta_{m}^{+}$ have the same definitions as in Eqs. (\ref{gamma}) and 
(\ref{eta}) [with $\tilde{v}_{0}({\bf R})$ replaced by the effective potential $\tilde{v}_{m}^{+}({\bf R})$], and are expressed at the reference point ${\bf R}_{0}$. In the derivation of Eq. (\ref{DquadraV0}), we have
used the relation $\eta_{m}^{+}({\bf R})=\eta_{m}^{+}+ \gamma_{m}^{+} \left[\tilde{v}_{m}^{+}({\bf R})-
\tilde{v}_{m}^{+}({\bf R}_{0})\right]$ which holds for any quadratic potential $\tilde{v}_{m}^{+}({\bf R})$.
We then go to Fourier space by writing
\begin{eqnarray}
f_{m;\lambda_{1};\lambda_{2}} (E) =\int d \tau F_{m;\lambda_{1};\lambda_{2}} (\tau) \, e^{- i E \tau},
\label{Fourier}
\end{eqnarray}
 and obtain a system of coupled first-order differential equations for $F$
\begin{eqnarray}
\left[
i(1+\gamma_{m}^{+} \tau^{2})
\frac{d }{d \tau}+i\gamma_{m}^{+} \tau -\eta_{m}^{+} \tau^{2}
+\omega-
\tilde{v}_{m}^{+}({\bf R}_{0})-E_{m,\lambda_{1}}
+i0^{+} \right]
F_{m;\lambda_{1};\lambda_{2}}(\tau) \nonumber \\
-\tilde{v}_{m}^{-}
F_{m;-\lambda_{1};\lambda_{2}}(\tau) = \delta(\tau) \, \delta_{\lambda_{1},\lambda_{2}}.
\label{systlast}
\end{eqnarray}
\end{widetext}
Introducing into Eqs. (\ref{systlast}) the (last) change in function,
\begin{eqnarray}
F_{m;\lambda_{1};\lambda_{2}}(\tau)= \frac{ h_{\lambda_{1};\lambda_{2}}
\left[t(\tau)\right]}{\sqrt{1+\gamma_{m}^{+} \tau^{2}}}
e^{i \left[\omega-\tilde{v}_{m}^{+}({\bf R}_{0}) \right]t(\tau)}
\nonumber \\
\times
e^{i\left( \eta_{m}^{+}/\gamma_{m}^{+} \right)
\left[t(\tau)- \tau \right]} \hspace*{0.5cm}
\end{eqnarray}
with
\begin{eqnarray}
t(\tau)=\frac{1}{\sqrt{\gamma_{m}^{+}}} \arctan\left( \sqrt{\gamma_{m}^{+}} \tau \right),
\end{eqnarray}
we arrive at a simple linear system of two coupled first-order inhomogeneous differential
equations with constant coefficients,
\begin{eqnarray}
\left[i \frac{d}{d t}-\lambda E_{m}+i0^{+} \right]
h_{\lambda;\lambda}(t)-\tilde{v}_{m}^{-} h_{-\lambda;\lambda}(t) &=&
\delta \left[\tau(t)\right], \hspace*{1cm} \label{s1}
 \\
\left[i \frac{d}{d t}+\lambda E_{m}+i0^{+} \right]
h_{-\lambda;\lambda}(t)-\tilde{v}_{m}^{-} h_{\lambda;\lambda} (t)&=& 0
\label{s2}
\end{eqnarray}
with
\begin{eqnarray}
\tau(t)=\frac{1}{\sqrt{\gamma_{m}^{+}}} \tan \left(\sqrt{\gamma_{m}^{+}}t \right).
\label{t}
\end{eqnarray}
Note that $ \delta \left[ \tau(t) \right]=\delta(t)$ if $\gamma_{m}^{+} \leq 0$ and
$\delta \left[ \tau(t) \right]=\sum_{n} \delta\left(t-n \pi/\sqrt{\gamma_{m}^{+}}\right)$ if $\gamma_{m}^{+} >0$.
Let us consider for the time being the case $\gamma_{m}^{+} \leq 0$. The solution of the system
of Eqs. (\ref{s1}) and (\ref{s2}) leading to a well-defined integral (\ref{Fourier}) can
then be readily derived and reads
\begin{eqnarray}
\left(
\begin{array}{c}
h_{\lambda;\lambda}(t) \\
h_{\lambda;-\lambda}(t)
\end{array}
\right)
= -\frac{i \theta(t)}{2} \left\{
\left(
\begin{array}{c}
1+\lambda \alpha_{m} \\
\beta_{m}
\end{array}
\right) e^{-it \sqrt{E_{m}^{2}+\left[\tilde{v}_{m}^{-} \right]^{2}}}
\nonumber \right. \\
\left.
+
\left(
\begin{array}{c}
1-\lambda \alpha_{m} \\
- \beta_{m}
\end{array}
\right) e^{it \sqrt{E_{m}^{2}+\left[\tilde{v}_{m}^{-} \right]^{2}}}
 \right\} e^{-0^{+}t}. \hspace*{0.5cm}
\end{eqnarray}
The expressions for the coefficients $\alpha_{m}$ and $\beta_{m}$ are given in
Eqs. (\ref{alpha}) and (\ref{beta}). Coming back to the original functions
$\tilde{g}_{m;\lambda_{1};\lambda_{2}}({\bf R})$ and setting ${\bf R}={\bf R}_{0}$, we get the compact expression
for the modified retarded Green's function written in Eq. (\ref{solution}). For
the case $\gamma_{m}^{+} >0$, it is important to realize that the relevant variable is
the time $t$, not the variable $\tau$ (whereas it is possible to work
indifferently with $t$ or $\tau$ for $\gamma_{m}^{+} \leq 0$). It can be checked that
integral (\ref{solution}) is well defined as well for $\gamma_{m}^{+} \leq 0$ as for
$\gamma_{m}^{+} >0$ [in the latter case the infinitesimal quantity $i0 ^{+}$ is crucial
while it does not help to make the integral convergent when expressing the
solution under the form of an integral over $\tau$ as within Eq.
(\ref{Fourier})].

\section{Solution for a locally quadratic mass term $V_{z}$ \label{B}}
\label{appB}
In this appendix, we solve the equations of motion, Eq. (\ref{Dysondiag}), in the regime where we can consider that the effective potential $\tilde{v}_{m}^{+}({\bf R})$ has a negligible spatial dispersion and that the effective
potential $\tilde{v}_{m}^{-}({\bf R})$ can be locally described up to its second-order derivatives. This regime contains as a particular case the situation where the scalar potential $V_{s}({\bf R})$ is globally constant in space and the mass potential $V_{z}({\bf R})$ has a quadratic dependence on ${\bf
R}$. In this particular case, we obviously get exactly that $\tilde{v}_{m}^{+}({\bf R})=cst$ and
$\tilde{v}_{m}^{-}({\bf R})$ depends quadratically on the variable ${\bf R}$,
\begin{eqnarray}
\tilde{v}_{m}^{+}({\bf R})
&=&
V_{s}({\bf R})-\frac{l_{B}^{2}}{4} \Delta_{{\bf R}} V_{z}({\bf R})=cst, \hspace*{0.6cm}
\\
\tilde{v}_{m}^{-}({\bf R})
&=& -
V_{z}({\bf R})-m \frac{l_{B}^{2}}{2} \Delta_{{\bf R}} V_{z}({\bf R}). \hspace*{0.6cm}
\end{eqnarray}
In the regime considered in this appendix, Eq. (\ref{Dysondiag}) becomes
\begin{widetext}
\begin{eqnarray}
\left(\omega- \tilde{v}_{m}^{+}-E_{m,\lambda_{1}}+i 0^{+} \right) \,
\tilde{g}_{m;\lambda_{1};\lambda_{2}}({\bf R})
=
 \delta_{\lambda_{1},\lambda_{2}}
+ \tilde{v}_{m}^{-}({\bf R}) \tilde{g}_{m;-\lambda_{1};\lambda_{2}}
({\bf R})+i\frac{l_{B}^{2}}{2} \left[ \partial_{X} \tilde{v}_{m}^{-} \partial_{Y}-
\partial_{Y} \tilde{v}_{m}^{-} \partial_{X}
\right] \tilde{g}_{m;-\lambda_{1};\lambda_{2}}({\bf R})
\nonumber
\\
-\frac{l_{B}^{4}}{8}
\left[ \left(
\partial_{Y}^{2}\tilde{v}_{m}^{-} \right) \partial_{X}^{2}
+
\left(
\partial_{X}^{2}\tilde{v}_{m}^{-}\right) \partial_{Y}^{2}
-2 \left(
\partial_{X} \partial_{Y}
\tilde{v}_{m}^{-}
 \right) \partial_{X}\partial_{Y}
\right]\tilde{g}_{m;-\lambda_{1};\lambda_{2}} ({\bf R}).
\label{DVzquadra1}
\end{eqnarray}
\end{widetext}

As in Appendix \ref{appA}, we introduce a reference point ${\bf R}_{0}$.
It can then be guessed that the functions
$\tilde{g}_{m;\lambda_{1};\lambda_{2}}({\bf R})$ are functionals of the
potential $E({\bf R})=\tilde{v}_{m}^{-}({\bf R}_{0})-\tilde{v}_{m}^{-}({\bf R})$, i.e., we can write
$\tilde{g}_{m;\lambda_{1};\lambda_{2}}({\bf R})
=f_{m;\lambda_{1};\lambda_{2}}\left[ E({\bf R})\right]$. The contributions
in Eq. (\ref{DVzquadra1}) involving the first-order derivatives of the function
$\tilde{v}_{m}^{-}({\bf R})$ then vanish. Furthermore, we shall suppose that the
equality (\ref{R1}) still holds in the present studied case, what can be
justified {\em a posteriori}. The problem thus reduces to the resolution of a
system of two coupled differential equations.
Indeed,
Eq. (\ref{DVzquadra1}) yields the system of equations
\begin{eqnarray}
\left[ (\gamma_{m}^{-} E +\eta_{m}^{-})\frac{d^{2}}{d E^{2}} +\gamma_{m}^{-} \frac{d}{dE}-
E+
\tilde{v}_{m}^{-}({\bf R}_{0})
\right]f_{m;-\lambda_{1};\lambda_{2}} (E) 
\nonumber
\\
-(\omega-\tilde{v}_{m}^{+}-E_{m,\lambda_{1}}+i0^{+})
f_{m;\lambda_{1};\lambda_{2}} (E)
=
 -\delta_{\lambda_{1},\lambda_{2}}
 \hspace*{0.5cm}
\label{DquadraVz}
\end{eqnarray}
with the coefficients $\gamma_{m}^{-}$ and $\eta_{m}^{-}$ expressed at the position ${\bf
R}_{0}$ and given by the formulas (\ref{gamma}) and (\ref{eta}) written for the potential
$-\tilde{v}_{m}^{-}({\bf R})$ in place of $\tilde{v}_{0}({\bf R})$. Applying the Fourier transformation, Eq. (\ref{Fourier}), we arrive at the
system,

\begin{eqnarray}
\left[
i(1+\gamma_{m}^{-} \tau^{2})
\frac{d }{d \tau}
+i\gamma_{m}^{-} \tau -\eta_{m}^{-} \tau^{2}+\tilde{v}_{m}^{-}({\bf R}_{0}) \right]
F_{m;-\lambda_{1};\lambda_{2}} (\tau)\nonumber 
\\
-(\omega-\tilde{v}_{m}^{+}
-E_{m,\lambda_{1}}+i 0^{+}
)
 F_{m;\lambda_{1};\lambda_{2}}(\tau) = - \delta(\tau) \,
\delta_{\lambda_{1},\lambda_{2}} \label{systVz}. \nonumber \\
\end{eqnarray}
Introducing into Eq. (\ref{systVz}) the change in function
\begin{eqnarray}
F_{m;\lambda_{1};\lambda_{2}}(\tau)= \frac{ h_{\lambda_{1};\lambda_{2}}
\left[s(\tau)\right]}{\sqrt{1+\gamma_{m}^{-} \tau^{2}}}
e^{i \left[\tilde{v}^{-}_{m}({\bf R}_{0})+\eta_{m}^{-}/\gamma_{m}^{-} \right]s(\tau)} \nonumber \\
\times e^{-i \left(\eta_{m}^{-}/\gamma_{m}^{-}\right) \tau }
\end{eqnarray}
with
\begin{eqnarray}
s(\tau)=\frac{1}{\sqrt{\gamma_{m}^{-}}} \arctan\left( \sqrt{\gamma_{m}^{-}} \tau \right),
\label{s}
\end{eqnarray}
a simpler system of two coupled differential equations with constant
coefficients comes out,
\begin{eqnarray}
i
\frac{d }{d s}
h_{-\lambda;\lambda} (s)
-(\omega-\tilde{v}_{m}^{+}
-\lambda E_{m}+i 0^{+}
)
 h_{\lambda;\lambda}(s)
\nonumber \hspace*{1.5cm} \\
=- \delta \left[ \tau(s)\right] , \hspace*{0.5cm}
\label{syst1} \\
i
\frac{d }{d s}
h_{\lambda;\lambda} (s)
-(\omega-\tilde{v}_{m}^{+}
+\lambda E_{m}+i 0^{+}
)
 h_{-\lambda;\lambda}(s)
=0
\label{syst2}. \hspace*{0.5cm}
\end{eqnarray}
Note that, in contrast to the situation encountered in Appendix \ref{appA}, the
variable $s$ does not have the meaning of the time here since it is no more
conjugated to the frequency $\omega$ [this is the reason why we took care of
naming the variable differently here although the expressions (\ref{t}) and
(\ref{s}) are almost identical]. After diagonalization of the $ 2 \times 2$ system, we
obtain that the homogeneous solution of Eqs. (\ref{syst1}) and (\ref{syst2}) is
\begin{eqnarray}
\left(
\begin{array}{c}
h_{\lambda;\lambda}(s) \\
h_{-\lambda;\lambda}(s)
\end{array}
\right)
=C \left(
\begin{array}{c}
\omega-\tilde{v}_{m}^{+}+\lambda E_{m} \\
-\kappa_{m}
\end{array}
\right) e^{i \kappa_{m} s} \nonumber \\
+ D \left(
\begin{array}{c}
\omega-\tilde{v}_{m}^{+}+\lambda E_{m} \\
 \kappa_{m}
\end{array}
\right) e^{-i \kappa_{m} s}
\end{eqnarray}
with $C$ and $D$ two arbitrary constants, and the energy $\kappa_{m}$ given by
Eqs. (\ref{kappa1}) and (\ref{kappa2}) (we forget for the time being the
infinitesimal quantity $i0^{+}$). The inhomogeneous solution of system of Eqs.
(\ref{syst1}) and (\ref{syst2}) is then obtained by varying the constants $C(s)$ and
$D(s)$. As a result, we get $C'(s)=-D'(s)=-i \delta(s)/(2\kappa_{m})$, that is
$C(s)=\mp i \theta(\pm s)/(2 \kappa_{m})$.
Using that
\begin{eqnarray}
\tilde{g}_{m;\lambda_{1};\lambda_{2}}({\bf R})= \int \! ds \frac{d\tau}{ds}
F_{m;\lambda_{1};\lambda_{2}}(\tau(s)) \, e^{i [\tilde{v}_{m}^{-}({\bf R}) -\tilde{v}_{m}^{-}({\bf R}_{0})]
\tau(s)} \label{ints},
 \nonumber \\
\end{eqnarray}
 the sign $\pm$ for the functions $C(s)$ and $D(s)$ is then chosen in such a way
that integral (\ref{ints}) is convergent with the help of the infinitesimal
quantity $i0^{+}$. Finally, taking the reference point ${\bf R}_{0}={\bf R}$ [so
that $E({\bf R})=0$], we arrive at the expression (\ref{gVzquadra}) for the
Green's function, which holds irrespective of the sign of the coefficient
$\gamma_{m}^{-}({\bf R})$.

\section{Simplifying the LDoS expression for locally flat potentials}
\label{appD}

In this appendix, we simplify further the expression (\ref{rhoFlat}) for the LDoS (valid for locally flat potentials) in the low-temperature regime within two different cases: (i) case of a potential landscape which varies slowly on the scale $L_{m}$ (Appendix \ref{appD1}); (ii) case of a potential landscape which fluctuates spatially in a random way on the scale $L_{m}$ (Appendix \ref{appD2}).

\subsection{Potentials flat on the scale $L_m$}
\label{appD1}

Writing the derivative of the Fermi-Dirac function as
\begin{eqnarray}
n_F'(\varepsilon-\xi_{m,\epsilon}({\bf R}))
= - \frac{1}{2} \int_{-\infty}^{+ \infty} \!\!\!\! \!\! dt \frac{Tt}{\sinh(\pi T t)} \, 
e^{ i t \left[ \varepsilon-\xi_{m,\epsilon}\left(
{\bf R}\right) \right]} ,
\nonumber
\\
\end{eqnarray}
and using the linearization of the effective energy 
$\xi_{m,\epsilon}({\bf R}+{\bf r})\simeq \xi_{m,\epsilon}({\bf r}) 
+{\bf R}.{\bm \nabla} \xi_{m,\epsilon}({\bf r})$ in Eq. (\ref{rhoFlat}), we can then perform the 
Gaussian integral over the vortex position ${\bf R}$ to get the LDoS expression,
\begin{widetext}
\begin{eqnarray}
\nonumber 
\rho^{STS}({\bf r},\varepsilon,T)
\simeq \frac{1}{2 \pi l_{B}^{2}} \frac{4}{\pi} \int_{-\infty}^{+ \infty} \!\!\!\! \!\! dt 
\left[ e^{it(\varepsilon-\xi_0({\bf r}))} 
\exp\left\{-\frac{t^2 \Gamma_0^{loc}({\bf r})^2}{4 A_s}\right\}
+ \frac{1}{2}\sum_{m=1}^{+ \infty} \sum_{\epsilon=\pm} 
\left\{ (1+\epsilon \beta_{m}) \frac{1}{m!}
\frac{\partial^{m}}{\partial s^{m}} \right.
\right. \nonumber \\
\left. \left. \left.
+ (1-\epsilon \beta_{m}) \frac{1}{(m-1)!}
\frac{\partial^{m-1}}{\partial s^{m-1}}
\right\}
\frac{1}{1-s}
 e^{it(\varepsilon-\xi_{m,\epsilon}({\bf r}))} 
\exp\left\{-\frac{t^2 \Gamma_{m,\epsilon}^{loc}({\bf r})^2}{4 A_s} \right\} 
\right|_{s=0} \right]
\label{dos1}
\end{eqnarray}
with $\Gamma_{m,\epsilon}^{loc}({\bf r})=l_{B} |{\bm \nabla} \xi_{m,\epsilon}({\bf r})|$
and assuming temperature is low enough [i.e., $T\ll\Gamma_{m,\epsilon}^{loc}
({\bf r})$] so that the limit $T\to 0$ can be taken.
We then perform the integral over time $t$, and obtain:
\begin{eqnarray}
\rho^{STS}({\bf r},\varepsilon,T) \simeq
\frac{1}{2 \pi l_{B}^{2}} \frac{4}{\sqrt{\pi}} 
\left[ \frac{1}{\Gamma_0^{loc}({\bf r})}
\exp\left\{-\left[\frac{\varepsilon-\xi_{0}({\bf r})}{\Gamma^{loc}({\bf r})} \right]^{2} \right\}
+ \frac{1}{2} \sum_{m=1}^{+ \infty} \sum_{\epsilon=\pm} 
\frac{1}{\Gamma_{m,\epsilon}^{loc}({\bf r})}
\left\{ (1+\epsilon \beta_{m}) \frac{1}{m!}
\frac{\partial^{m}}{\partial s^{m}} \right. \right.
\nonumber \\
\left. \left. \left.
+ (1-\epsilon \beta_{m}) \frac{1}{(m-1)!}
\frac{\partial^{m-1}}{\partial s^{m-1}}
\right\}
\frac{1}{\sqrt{1-s^{2}}}
 \exp\left\{-A_{s} \left(\frac{\varepsilon-\xi_{m,\epsilon}({\bf r})
}{\Gamma_{m,\epsilon}^{loc}({\bf r})}
\right)^{2} \right\} \right|_{s=0}
\right] . \hspace{0.5cm}
\label{dos2}
\end{eqnarray}
\end{widetext}
Finally, using the following relation \cite{Gradstein} obeyed by the Hermite 
polynomials $H_{n}(x)$ 
\begin{eqnarray}
\frac{1}{\sqrt{1-s^{2}}} \exp\left[ 
\frac{2s x^{2}}{1+s}
\right]=\sum_{n=0}^{+ \infty} \frac{(s/2)^{n}}{n!}\left[H_{n}(x)\right]^{2}, 
\end{eqnarray}
formula (\ref{dos2}) can be recast in expression (\ref{dosgradient}).

\subsection{Potentials random on the scale $L_m$}
\label{appD2}
We consider here the limit where the potential has strong spatial variations
along the cyclotron radius, which applies to the situation of large Landau
levels. We assume for simplicity that the antisymmetric part $V_z$ of the disorder 
potential can be neglected compared to the scalar component $V_s$
so that the effective potential given by Eq.~(\ref{effective}) reads
\begin{eqnarray}
\nonumber
\xi_{m,\pm}({\bf r}) 
&=& \pm E_m + \frac{1}{2} \int d^2 {\bm \eta} V_s({\bm \eta}) 
[ K_m({\bf R}-{\bm \eta})\\ 
&&\hspace{0.5cm} + K_{m-1}({{\bf R}-\bm \eta})],
\label{effectivescalar}
\end{eqnarray}
where $K_{-1}\equiv K_0$ in order for the above expression to apply
for $m=0$ as well.

The averaging procedure is carried through the isotropic distribution function 
$S(q)$ in Fourier space (here $q=|{\bf q}|$) that describes the spatial correlations of disorder
\begin{equation}
\overline{V_s({\bf R_1}) V_s({\bf R_2})} = S({\bf R_1}-{\bf R_2}) = 
\int \frac{d^2 {\bf q}}{(2\pi)^2} S(q) \, e^{i{\bf q}\cdot({\bf R_1}-{\bf R_2})}
\end{equation}
so that the spatially averaged LDoS becomes
\begin{eqnarray}
\nonumber
\rho^{DoS}(\omega) &\equiv& \overline{\rho({\bf r},\omega)} =
- \frac{4}{ \pi} \mathrm{Im} \int_{-\infty}^{+\infty} \!\!\! dt 
\int \!\!\!
\frac{d^{2}{\bf R}}{2 \pi l_{B}^{2}}
\sum_{m=0}^{+ \infty} \frac{1}{2} \sum_{\epsilon=\pm}\\
\nonumber
&& \hspace{-2cm}
\times  \int \mathcal{D}V_s \,
[K_{m}({\bf R}-{\bf r}) + K_{m-1}({\bf R}-{\bf r})]
e^{i[\omega-\xi_{m,\epsilon}({\bf R})] t}\\
&& \hspace{-2cm}  \times \exp\left\{-\frac{1}{2}\int d^2{\bf R_1} 
\int d^2{\bf R_2}\, S^{-1}({\bf R_1}-{\bf R_2}) 
V_s({\bf R_1}) V_s({\bf R_2}) \right\} \nonumber \\
&& \hspace{-2cm}
\end{eqnarray} 
where the distribution $S^{-1}$ obeys $\delta({\bf R})= \int d^2{\bm \eta}\; S^{-1}({\bf R}-{\bm \eta}) 
S({\bm \eta})$.
Inserting the effective potential, Eq.~(\ref{effectivescalar}), and
performing the functional integral over the disorder realizations,
we obtain
\begin{eqnarray}
\nonumber
\rho^{DoS}(\omega) &=& 
- \frac{4}{ \pi} \mathrm{Im} \int_{-\infty}^{+\infty} \!\!\! dt 
\int \!\!\!
\frac{d^{2}{\bf R}}{2 \pi l_{B}^{2}}
\sum_{m=0}^{+ \infty} \frac{1}{2} \sum_{\epsilon=\pm}
e^{i[\omega-\epsilon E_m] t}\\
&& \hspace{-1cm}
\times  
[K_{m}({\bf R}-{\bf r}) + K_{m-1}({\bf R}-{\bf r})] 
\exp \left\{ -\frac{1}{4} t^2\, [\Gamma_m^{DoS}]^2 \right\}
\nonumber \\ &&  \hspace{-1cm}
\end{eqnarray} 
where the width $\Gamma_m^{DoS}$ is given by Eq.~(\ref{width}).
The above expression has obviously become ${\bf r}$ independent
so that the ${\bf R}$ integral can be carried using the normalization
condition $\int d^2{\bf R} K_m({\bf R})=1$. The remaining time integral 
gives the final result quoted in Eq.~(\ref{average}).

\end{document}